# Machine Learning for Predicting Thermal Transport Properties of Solids

Xin Qian and Ronggui Yang*

School of Energy and Power Engineering,

Huazhong University of Science and Technology, Wuhan 430074, China

Email: ronggui@hust.edu.cn

ORCID: 0000-0002-3198-2014 (Xin Qian)

0000-0002-3602-6945 (Ronggui Yang)

**Abstract**

Quantitative descriptions of the structure-thermal property correlation have always been a challenging bottleneck in designing functional materials with superb thermal properties. In the past decade, the first-principles-based modeling of phonon properties using density functional theory and the Boltzmann transport equation has become a common practice for predicting the thermal conductivity of new materials. However, first-principles calculations of thermal properties are too costly for high-throughput material screening and multi-scale structural design. First-principles calculations also face several fundamental challenges in modeling thermal transport properties, for example, of crystalline materials with defects, of amorphous materials, and for materials at high temperatures. In the past five years or so, machine learning started to play a role in solving the aforementioned challenges. This review provides a comprehensive summary and discussion on the state-of-the-art, future opportunities, and the remaining challenges in implementing machine learning techniques for studying thermal conductivity. After a brief introduction to the working principles of machine learning algorithms and descriptors for characterizing material structures, recent research using machine learning to study nanoscale thermal transport is discussed. Three major applications of machine learning techniques for predicting thermal properties are discussed.

First, machine learning is applied to solve the challenges in modeling phonon transport of crystals with defects, in amorphous materials, and at high temperatures. In particular, machine learning is used to build high-fidelity interatomic potentials to bridge the gap between first-principles calculations and empirical molecular dynamics simulations. Second, machine learning can be used to study the correlation between thermal conductivity and other relevant properties for the high-throughput screening of functional materials. Finally, machine learning is a powerful tool for structural design to achieve target thermal conductance or thermal conductivity.

Keywords: Machine Learning, Thermal Properties, Thermal Conductivity, High Throughput Screening, Phonon, Molecular Dynamics, First Principles, Material Design

# 1. Introduction

Discovering and designing materials with superb thermal properties is of critical importance in a lot of technological applications. For example, the scaling of electronics has quickly approached the sub-10-nm regime, imposing ever more aggressive requirements on thermal management materials. The power electronics and optoelectronics industry is facing the challenge of dissipating high heat flux in the order of several 100 W/cm$^2$ or even above 1 kW/cm$^2$ while maintaining temperatures below 85 °C for silicon chips and ~ 100 °C for III-V and II-VI semiconductor chips [1, 2]. Furthermore, the heat dissipation on the chip is highly non-uniform because maximum chip heat flux, usually in the functional area, can be several times that of the surrounding region, resulting in hot spots in the devices. High thermal conductivity materials are therefore in urgent need as heat spreaders for removing hot spots. On the other hand, low thermal conductivity materials can find their applications when heat conduction needs to be minimized, such as thermoelectrics, thermal barrier coatings, and building envelopes. Reducing the thermal conductivity of thermal barrier coatings by half can lower the temperatures at the superalloy surface of the gas turbine blades by 55 K, and such high-temperature capability would be much more challenging to achieve only by increasing the melting point of superalloys [3]. Thermal insulation of building envelopes is essential for saving the energy consumption in heating and air-conditioning [4], which could even reach nearly half of the building's electricity bill in hot climates because the majority of the cooling load is due to the heat transmission across the building envelope [5]. There have been long endeavors in searching for materials with ultra-high and ultra-low thermal conductivities, and for functional materials with target thermal properties [6].

Understanding the structural-property correlation could provide the guidelines for the search of functional materials with superb thermal transport properties. One of the early attempts to capture the correlation between thermal conductivity $k$ and other material properties is by the Slack equation [7]:

$$k = \frac{A\bar{M}\delta\theta_D^3}{\gamma^2 T} N^{-\frac{2}{3}} \qquad (1)$$

where $A$ is a constant, $\bar{M}$ is the average atomic mass, $\delta$ is the characteristic bonding length whose cubic power $\delta^3$ is the mean volume occupied by each atom, $\theta_D$ is the Debye temperature, $\gamma$ is the Grüneisen parameter, $T$ is the absolute temperature, and $N$ is the number of atoms in the unit cell. Indeed, the Slack equation provides a qualitative guideline for searching materials with high or low thermal conductivity. For example, high thermal conductivity is usually found in stiffer materials with lighter atoms, because the mean group velocity is determined by $v = \sqrt{B\delta^3/\bar{M}}$, as shown in Figure 1a-b. On the other hand, low thermal conductivity materials usually contain heavy elements and weak interatomic bonding. Unfortunately, the scaling analysis based on Slack's equation is too simple and cannot be used for material screening, as one has to assume that some other properties stay the same across a wide range of materials. Correlation between thermal conductivity and other material properties could hardly be captured by a single scaling relation. Furthermore, thermal conductivity is usually not the only target property when searching for functional materials. For example, thermal barrier coatings require high stiffness yet a low thermal conductivity, while an ideal thermal interface material should possess high thermal conductivity but low mechanical stiffness, as shown in Figure 1b. Obtaining a unified framework for quantitative prediction of thermal transport properties from atomic structures is thus of great interest for designing multifunctional materials, especially for properties that are not easily predicted or sometimes even contradictory to our conventional wisdom.

Even after many decades of efforts, predicting thermal properties from material structures remained phenomenological, empirical, and rather qualitative than quantitative, until the development of *ab initio* or first-principles calculations for phonon properties in 2007 by using the density functional theory (DFT) and the Boltzmann transport equation (BTE) together [8]. Since then, *ab initio* phonon computation has been applied to model phonon transport in a wide range of semiconductors[8-14] and two-dimensional materials [15-26]. It has been shown that the predicted thermal conductivity has excellent agreement with measurement values in the literature [22, 23, 27, 28]. *Ab initio* phonon computation[29, 30] also leads to recent discoveries of extremely high thermal conductivity materials, such as cubic boron arsenide [31-33] and isotope-enriched boron nitride [34] with thermal conductivity values higher than 1000 W/mK. However, *ab initio* phonon calculations are still facing several challenges, including 1) modeling strongly anharmonic materials at high temperatures, especially for materials that might involve phase-change behavior [35-37], 2) high-fidelity modeling of phonon-defect interactions in nonperfect crystals such as alloys due to the questionable validity of the decades-old virtual crystal approximation [26, 38, 39], 3) predicting the thermal conductivity of amorphous materials in a first-principle manner when phonon gas model is no longer valid [40-42], and 4) performing high throughput material screening due to the high computational costs.

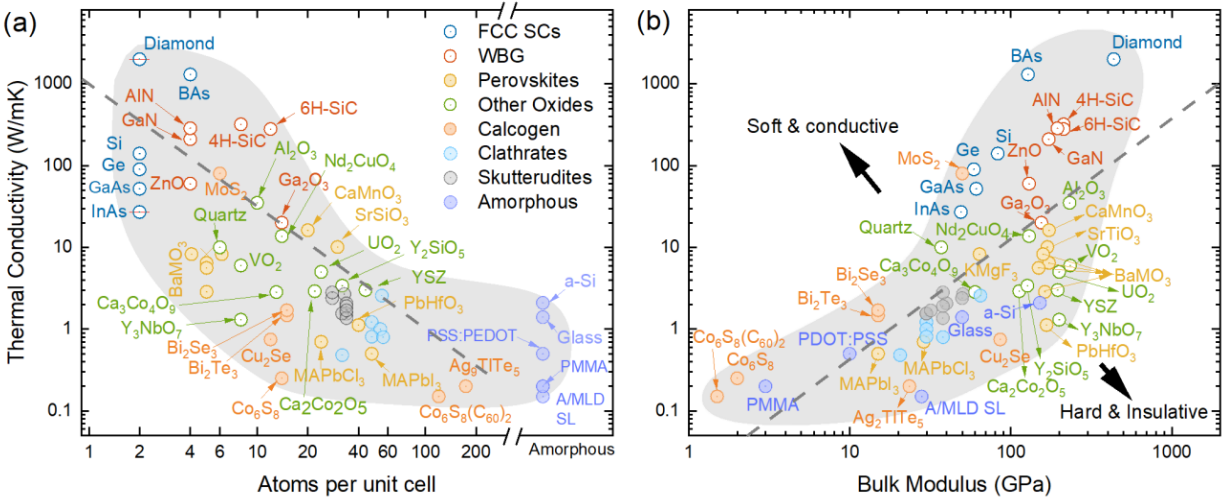

Figure 1. Correlation between the thermal conductivity of crystals and (a) number of atoms inside a unit cell and (b) the bulk modulus. The dashed line indicates the trend predicted by the Slack equation. Panel b is re-plotted based on ref. [43] with updated data points, Copyright 2018 by Wiley. Data is taken from Material Research Laboratory (MRL) Database [44].

In principle, the aforementioned challenges might eventually be solved through either development of physical models or numerical algorithms with considerable complexity, such as including higher-order phonon interactions for high-temperature materials. Either route requires long research and development cycles with limited efficiency. Machine learning (ML), on the other hand, can offer an efficient way to tackle these challenges by learning the embedded knowledge from data. Fundamental challenges in modeling thermal transport for materials at high temperatures, phonon-defect interactions, and thermal transport in amorphous materials can be solved by using ML algorithms. Specifically, ML methods are used to develop high-fidelity interatomic potentials for molecular dynamics (MD), which are known as machine learning potentials (MLPs). Since MD does not rely on the phonon gas model or the perturbation theory, and MD can easily incorporate detailed atomic structures, MD is the natural choice for modeling thermal transport in crystals with defects or in amorphous materials. MD also intrinsically incorporates temperature-dependent interatomic forces through direct tracking of vibrational trajectories of atoms. However, MD suffers from limited accuracy of empirical potentials using the "rigid" functional forms. Improving the accuracy of empirical interatomic potential is difficult, because the *ab-initio* potential energy surface (PES) in the atomic configurational space is high-dimensional, which can hardly be fitted by simple functional forms that are artificially assigned based on the pre-knowledge of the interatomic bonding [45-48]. In contrast, MLP can smartly fit the *ab initio* PES in a data-driven manner, without requiring "rigid" functional forms that can only fit specific shapes of the PES. MLP has been shown to have improved accuracy compared with empirical interatomic potential, which could be promising for modeling thermal properties.

ML could also play an important role in the high throughput screening of materials with target properties from millions of candidates. Development of material databases in recent years such as AFLOW [49], Materials Project [50], NOMAD [51], ICSD [52], and OQMD [53] has enabled researchers to perform high-throughput screening of materials by applying a series of screening criteria to narrow down promising candidate materials. For example, high throughput screening for high-efficiency thermoelectric materials usually rules out materials with positive formation enthalpy, dynamical instability, and zero electronic bandgaps [54, 55]. Such a pre-screening process could reduce the number of candidate materials from $\sim 10^5$ to $\sim 10^2$ [55]. Yet, due to the computational costs, it remains challenging or even unfeasible to perform *ab initio* phonon simulations to compute thermal conductivity for hundreds of candidate materials. Previously, empirical models like the Slack equation [7] or other semi-empirical models are used for high throughput predictions of thermal conductivity, but at the cost of accuracy or transferability [54, 56-59]. ML algorithms can serve as an alternative to performing high-throughput prediction of thermal conductivity with good accuracy, by learning the correlation between thermal conductivity and other properties such as chemical composition, lattice constants, atomic coordinates, bandgaps, sound velocities, Debye temperature, just to name a few. The idea is to use ML algorithms to train a surrogate model for making predictions of thermal conductivity from the other material properties that are available or easy to obtain. Such ML models can be orders of magnitude faster than the *ab initio* phonon calculations when dealing with the large pool of candidate materials.

In addition to high-throughput modeling and prediction of thermal properties, designing nanoscale material or device architectures with superior thermal transport would be crucial for both thermal management and energy conversion. However, identifying optimal structure with target property would be rather difficult, because multiple factors including structural parameters and physical properties could influence thermal transport, including length scales [60-62], grain boundaries [63], roughness [64, 65], vacancies or defects [26, 39], and others [66-68]. ML could play a role in the design of micro/nano-structures as it is particularly suitable for approximating unknown complex functions. In this case, the ML models are used in a "backward" manner: instead of predicting thermal properties with known structures, the goal is to identify optimal structure design with superior thermal properties in the entire possible design space.

This review discusses and summarizes how ML techniques have been applied to solve the aforementioned challenges in understanding thermal transport in solids. To begin with, the working principles of ML are introduced in Section 2, specifically on the databases, descriptors, and algorithms. In Section 3, recent applications of ML for thermal transport are summarized, in particular addressing challenges in modeling phonon-defect interactions and thermal transport in amorphous materials, predicting high-temperature thermal and vibrational properties, high throughput screening of optimal thermal conductivity, and structural design for optimal thermal conductance. Finally, limitations and challenges in applying ML techniques are discussed along with possible strategies to address them, including computational cost, large errors of extrapolation, data availability, and the iteration with experiments.

## 2. Fundamental Concepts of Machine Learning

The performance of ML algorithms to accomplish a certain task, such as imaging recognition, decision making, or regression, typically improve or optimize with "experience" encoded within the training dataset [69]. ML can be classified into three categories according to the tasks of the algorithms: supervised learning, unsupervised learning, and reinforcement learning [70]. Supervised learning tries to find the best approximation of an unknown function $\boldsymbol{y} = f(\boldsymbol{x})$ that maps the input variables ($\boldsymbol{x}$) to the target output variables ($\boldsymbol{y}$). If the unknown function $f$ is continuous, then approximating this unknown function is essentially a regression problem. Regression is performed using a database that samples the landscape of the function $f$. The other type of problem in supervised learning is the classification problem such as recognizing images of handwritten digits, which can also be regarded as approximating a discretized function [71]. Unsupervised learning is concerned with looking for patterns and data clusters in a dataset, in which the data entries in the dataset are not labeled as input or output variables. The unsupervised learning algorithm can discover the clustering of data without any human intervention. One good example is to discover phase diagrams from the spin states of an Ising lattice [72]. The temperature and spin states of the Ising lattice are both input variables in the database without any variables labeled as target outputs, but the spin states belonging to the same phase are regarded as belonging to the same data cluster by the unsupervised learning algorithm. Finally, reinforcement learning is a computational approach seeking to take action through interactions with a reward function. The learner is not instructed on which actions to take but instead must discover the actions yielding the maximum value of the reward function by trying them. For example, reinforcement learning methods were used for training computers for playing the Go game, where the ML program must decide the moves yielding a maximum probability of winning the game [73]. By far the most widely used ML methods in material science and thermal science belong to supervised learning, which will therefore be the focus of this review.

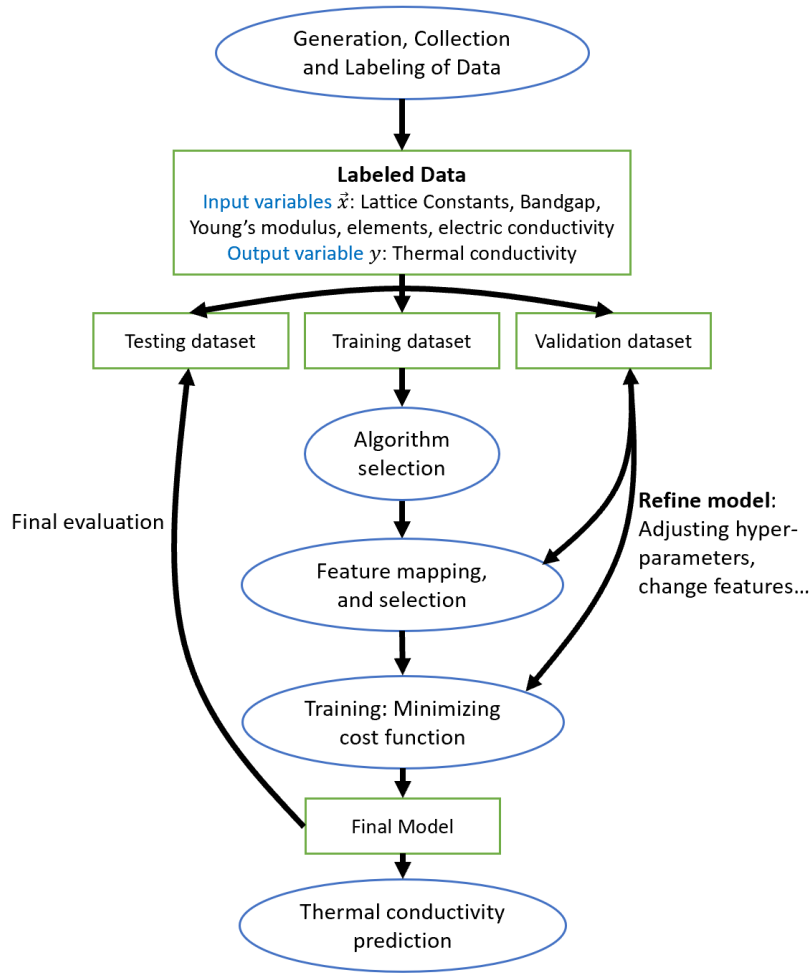

Figure 2. Workflow of supervised learning for thermal conductivity prediction.

One grand challenge in thermal science is to extract a quantitative correlation for thermal conductivity with material structure and other properties, such as chemical compositions, lattice constants, elastic constants, and so on. This is indeed a typical regression problem that can be solved using supervised learning algorithms to approximate the unknown function or mapping $k = f(\mathcal{M})$, where $\mathcal{M}$ labels the candidate material in the material design space to the thermal conductivity $k$. Here we will use this problem as an example for explaining the workflow of supervised learning, which is summarized in Figure 2. In general, supervised learning begins with generating or collecting data to build a database. The data entries in the database should be labeled as input variables and target output variables. Here, the thermal conductivity data is labeled as target output variables while all other material properties or structures are labeled as input variables. Then the database is separated into three datasets for training, testing, and validation. The training dataset is used for training the supervised learning model until the desired accuracy of reproducing the output variables in the training dataset is achieved. Algorithms should be selected based on the complexity of the problem and the size of the accessible training dataset, which will be discussed in more detail later. After selecting the algorithm, a key ingredient for supervised learning is to translate the raw input variables $x$ into descriptors $q$ with suitable math forms as input for the ML algorithms. Components of the descriptor vector $q$ are called features and the process translating the input variables to descriptors is usually referred to as feature extraction. The simplest examples of feature extraction include: using atomic numbers and masses to represent the elements in the materials, and using binary arrays [0101101…] to represent the distribution of vacancies or substitution defects in a crystal. It is also important to realize that not all features or material properties are relevant or useful to predict the thermal conductivity of solids. For example, electrical conductivity has a weak correlation with thermal conductivity for semiconductors and dielectric materials. Including irrelevant features would increase the dimensionality of the descriptors and would result in bad performance or slow training of the ML models, which is known as the "curse of dimensionality" [74]. Therefore, feature selection is crucial for removing irrelevant features to reduce the complexity and dimensionality of the regression problems. After feature selection, the ML model can be trained by optimizing its performance, usually through minimizing by some cost function (e.g. squared difference between the prediction $f(\bm{q}_i)$ and reference thermal conductivity data $k_i$, *i.e.* $\sum_i |k_i - f(\bm{q}_i)|^2$). The training process involves the adjustment of parameters that control the learning rate and accuracy of the ML model. During the training process, it is important to consider issues of underfitting and overfitting, and a validation dataset should be used to evaluate whether a model is underfitting or overfitting. Underfitting simply means that the ML model is not versatile enough to capture the complexity of the unknown function $f$, hence there will be a high bias (large error) when the model is evaluated using the validation dataset. Underfitting can be solved by including more adjustable parameters. On the other hand, overfitting is caused by the presence of too many adjustable parameters compared with the size of the training dataset. An overfitted model could reproduce the functional values in the training dataset very accurately but tends to regard the noises in the training dataset as relevant information. As a result, an overfitted model could fail to make reliable predictions for the data points beyond the training dataset. It is therefore recommended to stop the training process to avoid overfitting, once the validation error stops decreasing. Finally, the testing dataset is used to evaluate the regression accuracy. Figure 2 shows how the trained model can be used to predict thermal conductivity from the material structures.

The remaining part of this section will focus on more detailed discussions on key ingredients of supervised learning: datasets, descriptors, and supervised learning algorithms. Section 2.1 focuses on datasets of material properties and we will address the limited data accessibility for studying thermal properties. In Section 2.2 we will discuss the necessity and working principles of descriptors for thermal conductivity prediction. In Section 2.3, several supervised learning algorithms with their pros and cons will be discussed. More specifically, kernel-based learning algorithms rooted in the theory of linear functional spaces are first summarized in Section 2.3.1, and then nonlinear models such as decision-tree-based models and neural networks will be covered in Section 2.3.2.

## 2.1. Datasets

The size and quality of the datasets lay the foundation for the success of supervised learning. For studying thermal transport properties, the data can be classified as the following types: (I) general properties of the crystal, such as density, specific heat, lattice constants, formation enthalpy, electronic bandgap, elastic modulus, and the associated elemental properties such as atomic mass, radius, valance states, *etc.*, (II) properties of interatomic bonds, including the potential energy and interatomic forces, (III) harmonic properties such as harmonic force constants, phonon dispersion, group velocities, vibrational density of states, and the scattering phase space, and (IV) anharmonic properties such as third- and higher-order force constants, Grüneisen parameters, phonon lifetimes and mean free paths. Note that the scattering phase space is defined as the volume of the Brillouin zone where phonon scatterings satisfying the energy and momentum conservation can happen.[75] Although phonon scatterings are anharmonic processes, scattering phase space can be calculated only using harmonic phonon properties and therefore belongs to type III. Thermal conductivity also belongs to property type I but is usually labeled as the target output variable. Depending on the implementation of ML for thermal property predictions, databases are established in different ways. For example, along the line of developing MLP for calculating thermal and phonon properties using molecular dynamics, the database essentially samples the landscape of the interatomic potential energy surface (type II data), which are usually generated by performing DFT simulations for different atomic configurations. Instead, for high-throughput screening and prediction of thermal conductivity, the goal is to directly correlate thermal conductivity to other properties for a wide range of materials, which heavily relies on the availability of type I data. The recent development of online databases such as AFLOW [49], Materials Project [50], NOMAD [51], ICSD [52], and OQMD [53] have collected various Type I material properties for thousands of materials, making it possible for high-throughput prediction and for accelerating material discovery. However, these material properties might not have a strong correlation with thermal conductivity, which could compromise the accuracy of the ML model. One way to mitigate this problem is to focus on harmonic vibrational properties (type III) and anharmonic vibrational properties (type IV) that are directly used to calculate thermal conductivity. Unfortunately, the availability of type III and IV data is rare, due to the computational complexities of phonon calculations. Recently, a database for harmonic phonon properties was built by Togo *et al.* for ~$10^4$ crystals, including semiconductors Si, Ge, GaN, BAs, layered materials $MoS_2$, $MoSe_2$, h-BN, oxides ZnO, $Al_2O_3$, $Ga_2O_3$, $HfO_2$, $SrTiO_3$, *etc.* [76]. AlmaBTE provided a small database containing anharmonic force constants of 58 semiconductors [77], which is an important step but the size is quite limited for training ML models. Due to the lack of data, it is pivotal to choose appropriate descriptors and ML algorithms with good performance using small datasets to prevent large errors and to avoid overfitting.

## 2.2. Descriptors

Representing data in appropriate mathematical forms is critical for the performance of ML models. In general, material descriptors can contain a wide range of information, which can be categorized into elemental representations, structural configurations, and material properties.

Elemental representations include physical quantities of the atomic species such as atomic numbers, ionization energy, electronegativity, ionic radius, and so on. Structural representations include lattice parameters, space groups, atomic coordination, X-ray diffraction patterns, and angular or radial distribution functions. Finally, examples of material properties that can be included in descriptors are Young's modulus, electronic bandgaps, sound velocities, Grüneisen parameters, *etc*. The complexity of constructing descriptors could vary a lot depending on the purposes. For example, for developing MLP, the descriptors are required to resolve the structural changes due to atomic vibrations from the equilibrium structure. On the other hand, for high throughput screening of thermal conductivity, it is no longer necessary to resolve atomic displacements, but the goal is to select the appropriate combination of elemental, structural representations, and some relevant material properties.

The representations of atomic configurations are the most intensively studied descriptors in material science, which are used to develop machine learning potentials (MLPs). In general, any interatomic potentials can be viewed as the mapping: $(E, \boldsymbol{F}) = f(\mathcal{X})$, which takes an atomic configuration $\mathcal{X}$ as input and predicts the associated energy $E$ and forces $\boldsymbol{F}$. Instead of using empirical functional forms, MLP utilizes machine learning regression algorithms to smartly approximate such mapping $f$. Developing MLP begins with feature extraction, namely representing atomic configurations $\mathcal{X}$ with appropriate mathematical forms, namely descriptors. The most straightforward choice of material descriptor for atomic configuration is the list of Cartesian coordinates for all atoms. Unfortunately, Cartesian coordinates fail to uniquely fingerprint the atomic structure. With any permutation, bulk translation, or rotation, a different list of coordinates can be generated corresponding to the same atomic structure, as shown in Figure 3a. Cartesian coordinates are an *overcomplete* set of descriptors for atomic configurations, meaning that the descriptor itself is unambiguous, but there exist multiple different descriptors corresponding to the same structure, as shown in Figure 3b. Overcomplete descriptors would result in low training efficiency because an extra amount of data are required to teach the ML model. Therefore, an essential requirement for descriptors of atomic structures is being invariant to the symmetry operations such as permutation, bulk rotation, and translations. On the other hand, *incomplete* descriptors would fail to distinguish between different atomic configurations, which would significantly compromise the MLP accuracy. One example of the incomplete descriptor is shown in Figure 3c proposed by Pozdnyakov *et al.* [78]. Consider four different types of atoms that are distributed around a circle surrounding the atom at the center. One possible way to represent such atomic configuration is through the histogram of angles for triplets of atoms. The two atomic configurations shown in Figure 3c are distinct but they have the same histogram of atomic angles. Such an incomplete set of descriptors is ambiguous which could map different atomic structures to the identical descriptor as shown in Figure 3d. This leads to low accuracy of regression. Ideally, a descriptor for the atomic configurations should be *complete*, meaning that the mapping $q = \mathcal{F}(\mathcal{X})$ between the descriptors $q$ and the atomic configurations $\mathcal{X}$ should have one-to-one correspondence (i.e. mapping $\mathcal{F}$ is a bijection) as shown in Figure 3e-f. One of the most successful and commonly used methods to construct a complete set of many-body descriptors is referred to as the Smooth Overlap of Atomic Positions (SOAP). The SOAP descriptor starts with constructing local atomic density function:

$$\rho_i(\boldsymbol{r}) = \sum_j \delta(\boldsymbol{r} - \boldsymbol{r}_j) \cdot f_{cut}(r_{ij}) \sim \sum_j \exp\left(-\frac{|\boldsymbol{r} - \boldsymbol{r}_j|^2}{\sigma_a^2}\right) f_{cut}(r_{ij}) \qquad (2)$$

where $j$ sums over all neighboring atoms of $i$ within the cutoff radius, and $f_{cut}(r_{ij})$ is a cutoff function that vanishes to zero when the interatomic distance $r_{ij} = |r_j - r_i|$ is greater than a certain cutoff radius. The summation over $j$ ensures permutation invariance ($\rho_i$ remains identical by any two neighboring atoms), and taking the relative position $r - r_j$ ensures the translational invariance ($\rho_i(r + T) = \rho_i(r)$ for any bulk translation $T$). Instead of treating the atoms as points which are represented as Dirac delta functions in the atomic density, SOAP smoothes the delta functions to a Gaussian function with a certain width $\sigma_a$, as shown in Figure 3e. Such smoothing over atomic density would result in better numerical behavior of descriptor vectors since delta functions show strong numerical changes in descriptor even with slight deviations of atomic positions from the equilibrium. The local atomic density $\rho_i(r)$ is then projected to a set of radial basis functions $g_n(r)$ and spherical harmonics $Y_{lm}(\phi, \theta)$, and the descriptor vector $q$ is finally obtained by computing the power spectrum:

$$\rho_i(r) = \sum_{nlm} c^i_{nlm} \cdot g_n(r) Y_{lm}(\phi, \theta) \tag{3}$$

$$(\vec{q}_i)_{nn'l} = \sum_m (c^i_{nlm})^* c^i_{nlm} \tag{4}$$

One limitation of the SOAP descriptor is that the computational cost increases dramatically with the number of atomic species in the material. To solve this problem, another complete set of symmetry-invariant descriptors called the "moment tensor" were developed based on the atomic distribution in the surrounding of an atom [79]. Instead of using atomic density distribution $\rho_i(r)$, the idea is to construct moment tensor of neighboring atoms to represent the chemical environment around a certain atom $i$:

$$M_{\mu\nu} = \sum_j f_\mu(|\boldsymbol{r}_{ij}|) \cdot \underbrace{\boldsymbol{r}_{ij} \otimes \ldots \otimes \boldsymbol{r}_{ij}}_{\nu \text{ times}} \quad (5)$$

where $f_\mu$ is a radially decaying function with the vector $\boldsymbol{r}_{ij} = \boldsymbol{r}_j - \boldsymbol{r}_i$ is the relative position of the neighboring atom $j$ with respective to atom $i$, and $\otimes$ denotes the tensor product. Such moment tensor showed comparable accuracy for building MLP when compared with SOAP, but with much higher computational efficiency [80]. There also exist other descriptors for atomic configurations, such as atom-centered radial and angular symmetry functions [81], Weyl matrices [82], crystal graphs [83], and Coulomb matrices [84], and atomic orbital matrices [85], while most of these might suffer from the problems of incompleteness or overcompleteness. A comprehensive discussion on the completeness of descriptors for atomic structures is published recently [78]. It should be noted that in some cases, despite that the descriptor itself might not form a complete representation, the problem can be mitigated by using some ML algorithms like convolution neural networks that intrinsically perform further feature extraction or selection [83].

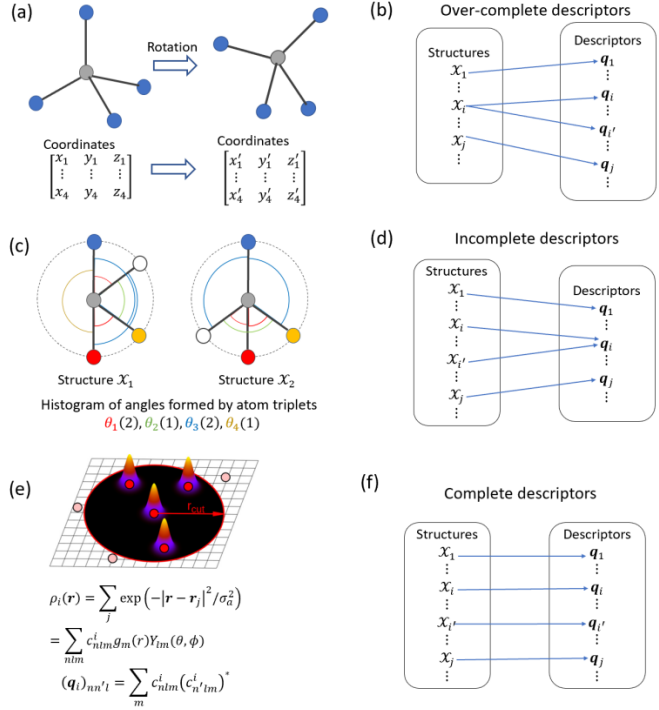

Figure 3. Completeness of atomic structure descriptors. (a) Rotation of an atomic structure would generate a new list of Cartesian coordinates, using CH$_4$ molecule as an example of the overcomplete descriptor, and (b) Mapping structure of overcomplete descriptors. (c) An example of an incomplete descriptor, where four different types of atoms are distributed around a circle of a centering atom. The descriptor is constructed through the histograms of different atomic angles formed by atomic triplets. Arcs of different colors indicate different atomic angles. The two atomic clusters have a distinctly different structure, but they correspond to the same histogram of atomic angles. Reproduced from ref. [78], Copyright 2020 by American Physical Society. (d) Mapping structure of incomplete descriptors. Reproduced from ref. [86], Copyright 2019 by American Physical Society. (e) A schematic of constructing smooth over atomic position (SOAP) descriptor, which is complete. (f) Mapping structure of complete descriptors.

For high-throughput prediction and screening of materials with target thermal properties, the descriptor needs to include a much wider range of information compared with descriptors for MLPs. Such descriptors should contain information such as atomic numbers, atomic masses and radius, lattice constants, coordination numbers, and so on. One typical way of constructing descriptor vectors is to include all elemental or structural properties into a descriptor vector is presented by Seko *et al*. [87] as shown in Figure 4. Considering a crystal with $N_a$ atoms in the unit cell, and each atom represented by $N_{ele}$ elemental properties and $N_{st}$ structural properties, the representation matrix can then by constructed as:

$$X = \begin{bmatrix} x_1^1 & x_2^1 & \cdots & x_{N_x}^1 \\ x_1^2 & x_2^2 & \cdots & x_{N_x}^2 \\ \vdots & \vdots & \ddots & \vdots \\ x_1^{N_a} & x_2^{N_a} & \cdots & x_{N_x}^{N_a} \end{bmatrix} = \begin{bmatrix} x^1 \\ x^2 \\ \vdots \\ x^{N_a} \end{bmatrix} \quad (6)$$

where $x_j^i$ is the $j$-th descriptor entry of atom $i$ in the unit cell, $x^i$ is the $i$-th row vector, and $N = N_{ele} + N_{st}$ is the total number of features for each atom. Each row-vector $x^i$ of the matrix $X$ can be viewed as a data point in a $N_x$-dimensional feature space, and the representation matrix $X$ can therefore be regarded as a probability distribution in the feature space. Therefore, one can construct the descriptor by looking at the statistical observables such as the mean and standard deviation of the row vectors. A simple but useful way to construct descriptor $q$ is through the mean value of the scattered points $x^i$ in the descriptor space:

$$q = \frac{1}{N_a} \sum_i x^i \quad (7)$$

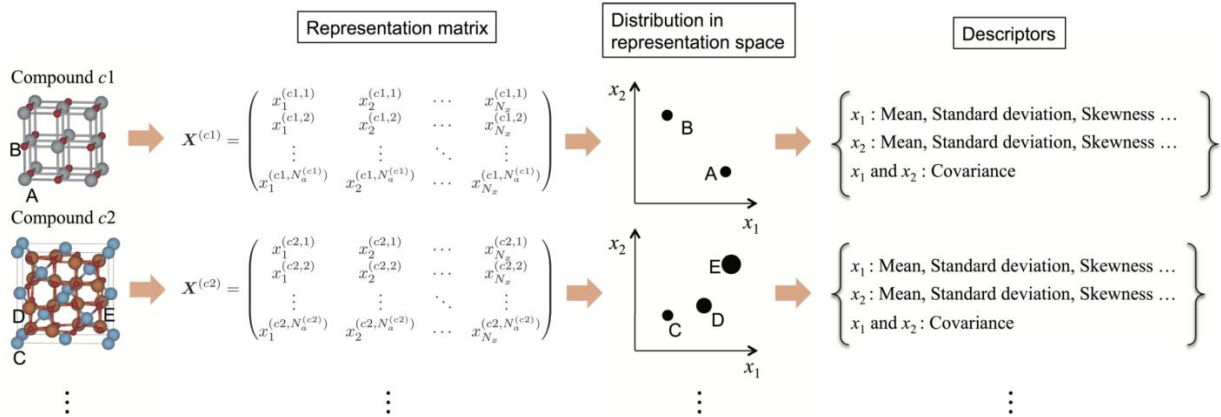

Figure 4. Schematic describing the construction of descriptors. Representation matrices collecting the structural and elemental properties are regarded as a probabilistic distribution in a representation space, and the statistical quantities such as mean, standard deviations, and covariances are derived as descriptors. Reproduced from ref. [87], Copyright 2017 by American Physical Society (APS).

The most important concern in developing descriptors for high throughput screening of thermal conductivity is to determine which properties are significant among a wide range of elemental, structural representations and properties. Among the available material properties, there could be a lot of redundant or irrelevant features. Irrelevant features might exist due to the weak correlations between the target property and certain input variables, such as the thermal conductivity and electric conductivity of insulators. It is also possible that some groups of features are all relevant to the output, but they are inter-correlated. For example, density can be calculated from atomic masses and lattice constants. Including all these three parameters in the descriptor does not provide new information than two of those. Including redundant and irrelevant features into the descriptors could lead to increased training costs or even compromised accuracy. Feature selection is therefore crucial for high-throughput screening and prediction of thermal properties. Feature selection can be performed either based on physical knowledge of thermal transport, or through feature selection algorithms, which will be discussed briefly in Section 3.2.

## 2.3. Supervised learning algorithms

To illustrate the supervised learning algorithms, let us now consider a set of materials $\mathcal{M}_1, \mathcal{M}_2, \ldots, \mathcal{M}_N$, such as a set of binary alloys with $N$ different compositions, and they are fingerprinted by the descriptor vectors $\boldsymbol{q}_1, \boldsymbol{q}_2, \ldots, \boldsymbol{q}_N$. This section focuses on how supervised learning algorithms work to approximate the unknown function $y = f(\boldsymbol{q})$, where the output variable $y$ could be energy or interatomic forces when developing interatomic potential, or thermal conductivity itself for high-throughput screening purposes. The performance of the algorithms would be affected by the size of dataset $N$, and the dimensionality of the descriptor vector $\boldsymbol{q}$. Generally, simple algorithms based on the theory of linear functional space and kernel functions have fair performances for small datasets that have only hundreds or thousands of data entries, but their scaling behavior of computational cost with data size $N$ limits their application to big datasets with millions of data entries. In contrast, nonlinear models usually perform better for big datasets with the high dimensionality of descriptors, while they might overfit when the dataset is too small. It is therefore necessary to have a basic understanding of how different algorithms can "learn" from data. To begin with, algorithms based on kernel functions will first be introduced in Section 2.3.1. Section 2.3.2 then focuses on the two nonlinear algorithms often used in the thermal science community, artificial neural networks and decision tree-based random forests. In addition, convolution neural networks and random forests can also be used for feature selection. This section will briefly discuss the working principles of these algorithms and compare their performances and limitations, while more detailed discussions can be found in refs. [88, 89].

### 2.3.1. Kernel-based learning algorithms

Kernel-based methods usually involve specifying kernel functions as a measure of similarity between the data points in the database, and the kernel functions map a nonlinear regression problem to a linear regression problem in a functional space with higher dimensionality. To begin with, let us consider an example of developing machine learning potentials (MLPs) for calculating thermal conductivity using MD simulations. The MLP is trained by fitting the *ab initio* potential energy surface. The most straightforward way to solve this regression problem is by minimizing the loss function:

$$L(\boldsymbol{w}) = \sum_i \|\boldsymbol{w}^T \boldsymbol{q}_i - y_i\|^2 + \lambda \|\boldsymbol{w}\|^2 \tag{8}$$

where $\boldsymbol{w}$ are the linear coefficients, $y_i$ represents output variables such as potential energy and interatomic forces for developing MLPs, with the subscript $i$ represent the index of the atomic configuration. $\boldsymbol{q}_i$ denotes the descriptor vector for the atomic configuration $i$. Compared with linear regression, there is an extra regularization term $\lambda \|\boldsymbol{w}\|^2$ in the loss function. Such a regularization term is especially useful for suppressing the noises and uncertainties in energy and forces that originated from finite thresholds of convergence or characteristics of the pseudopotentials in DFT simulations. The noise-suppression feature is due to the regularization term $\lambda \|\boldsymbol{w}\|^2$ that favors small absolute values of the linear coefficients. As a result, even if some noisy data points are lying far away from the line $y = \boldsymbol{w}^T \boldsymbol{q}$, the regularization prevents these outliers from changing the $\boldsymbol{w}$ dramatically during the training. For larger values of $\lambda$, the linear ridge regression has a stronger tendency to neglect noise in the dataset (smaller variance), but it would compromise the fitting accuracy (larger bias), which is referred to as the variance-bias tradeoff [88]. However, linear ridge regression might be too simple for fitting the high-dimensional potential energy surfaces, because there is no guarantee that the function $f(\boldsymbol{q})$ corresponding to the energy surface will be linear in the feature space of descriptors. To deal with the nonlinear problem, the kernel ridge regression (KRR) method introduces a set of basis functions $\boldsymbol{\phi} = [\phi_1, \phi_2, \ldots, \phi_M]^T$, which maps the data point $\boldsymbol{q}$ to higher dimensional functional space, where the original nonlinear regression problem is converted into a linear regression problem as shown in Figure 5. Here, the dimensionality $M$ of the basis function is usually much larger than the dimensionality of the descriptor $N$, which ensures that the model is flexible enough to capture nonlinearities. In some cases, the dimensionality $M$ could even approach infinity, such as the basis functions corresponding to the Gaussian kernels, which will be introduced later. Through decomposing the nonlinear energy surface into the basis set $f(\boldsymbol{q}) = \boldsymbol{w}^T \boldsymbol{\phi}(\boldsymbol{q})$, the loss function for nonlinear regression is then simplified by replacing $\boldsymbol{q}_i$ in Eq. (8) to $\boldsymbol{\phi}(\boldsymbol{q}_i)$:

$$L(\boldsymbol{w}) = \sum_j \left\| \boldsymbol{w}^T \boldsymbol{\phi}(\boldsymbol{q}_j) - y_j \right\|^2 + \lambda \|\boldsymbol{w}\|^2 \qquad (9)$$

Although one can obtain the coefficients $\boldsymbol{w}$ after carefully selecting the orthonormal basis functions $\boldsymbol{\phi}$, the implementation of the algorithm can be much simplified by expressing $\boldsymbol{w}$ as a linear combination of the basis functions $\boldsymbol{w} = \sum_i \alpha_i \boldsymbol{\phi}(\boldsymbol{q}_i)$, where the basis set $\boldsymbol{\phi}(\boldsymbol{q}_j) = [\phi_1(\boldsymbol{q}_j), \phi_2(\boldsymbol{q}_j), \ldots]^T$ and the coefficients $\alpha_i$ become the unknown parameters to be determined through the training process. It turns out that through such treatment, one never needs to specify the basis set $\boldsymbol{\phi}$, but only one kernel function $K(\boldsymbol{q}_i, \boldsymbol{q}_j) = \boldsymbol{\phi}^T(\boldsymbol{q}_i) \boldsymbol{\phi}(\boldsymbol{q}_j)$ expressed as the inner product of the basis functions. The solution to minimizing the $L(\boldsymbol{w})$ is solely determined by the kernel function $K$, which can be expressed in the matrix form as:

$$\boldsymbol{\alpha} = (\boldsymbol{K} + \lambda \boldsymbol{I}_N)^{-1} \boldsymbol{y} \qquad (10)$$

where elements of the $K$ matrices are $K_{ij} = K(q_i, q_j)$, $I_N$ is an N-by-N identity matrix, and $\alpha = [\alpha_1, \alpha_2, ..., \alpha_i, ...]^T$. Similarly, the unknown function $f$ is simplified to a linear superposition of kernel functions as well:

$$f(q) = w^T \phi(q) = \sum_i \alpha_i \phi^T(q_i) \phi(q_j) = \sum_i \alpha_i K(q_i, q) \tag{11}$$

After obtaining $\alpha$, unknown function $f$ can be computed using Eq. (11) at any point $q$ in the material design space.

These kernel functions can be interpreted as the measure of cosine similarity in the kernel space between atomic configurations $i$ and $j$ in the dataset. Eq. (11) essentially describes the prediction process of the kernel method. For an unknown atomic configuration $q$, the kernel-based model first compare it with every atomic configuration in the training dataset and evaluate the similarity value $K(q_i, q)$. The training data points will then collectively give the prediction value through the linear combination of kernel functions. The atomic structure most similar to $q$ will have the largest similarity value and thereby tends to have the strongest influence on the prediction result $f(q)$. A common choice of kernel function has a Gaussian form:

$$K(q, q') = \exp\left[-\frac{\|q - q'\|^2}{2\sigma_l^2}\right] \tag{12}$$

The Gaussian kernels correspond to a transformation $\phi$ with infinite dimensionality. Such definition of kernel functions guarantees that the similarity measure is the maximum when the two atomic configurations $q$ and $q'$ are identical, and decays to zero when the configurations $q$ and $q'$ are very dissimilar, i.e. $\|q - q'\| \to \infty$. $\sigma_l$ here is a hyperparameter that determines the decaying of correlation between different configurations. KRR is one of the most used algorithms in materials science when the dataset is small due to its simplicity. However, its performance is fundamentally limited by the variance-bias tradeoff and the selection of the parameter $\lambda$.

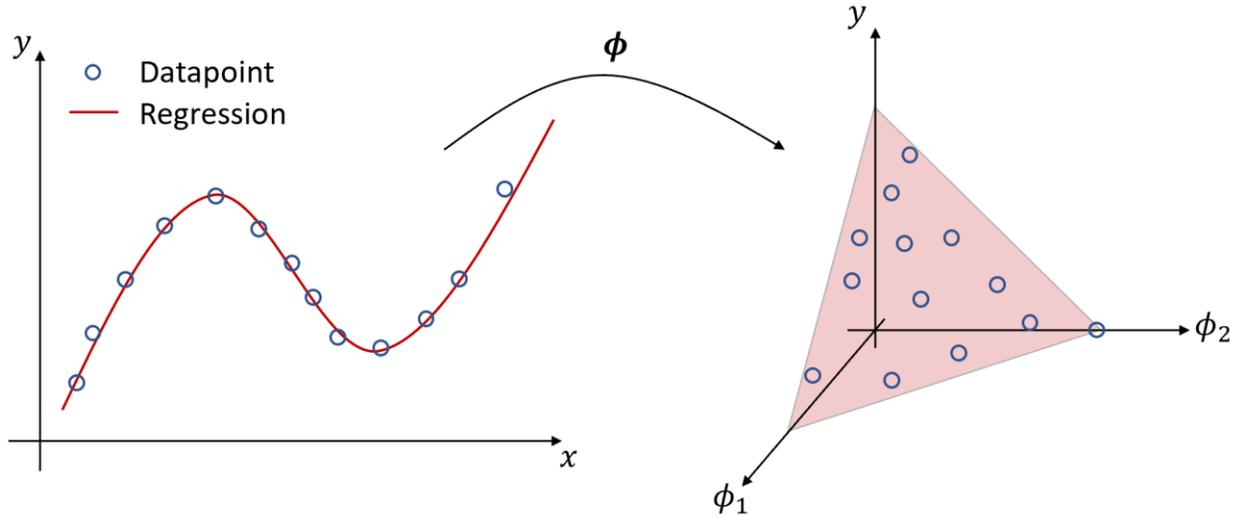

Figure 5. Transformation $\boldsymbol{\phi}$ converts a nonlinear regression problem to a linear regression problem in the higher dimensional space. The data points in higher dimensional space spanned by $\boldsymbol{\phi}$ can be fitted by a hyperplane.

Other kernel-based algorithms utilize the similar idea of mapping a nonlinear regression problem to a linear problem in a higher-dimensional space, but the loss functions are defined differently. For example, another commonly used kernel-based algorithm is the support vector machine (SVM) [90]. SVM is originally designed for data classification by drawing a classification boundary that maximizes the margin between the boundary and its closest data points. SVM can be used to solve the regression problems for thermal conductivity prediction. This is achieved by minimizing the so-called $\varepsilon -$ insensitive loss function:

$$L_\varepsilon = \begin{cases} 0, & \left| y - \sum_i \alpha_i K(\boldsymbol{q}_i, \boldsymbol{q}) - b \right| < \varepsilon \\ \left| y - \sum_i \alpha_i K(\boldsymbol{q}_i, \boldsymbol{q}) - b \right| - \varepsilon, & \text{otherwise} \end{cases} \quad (13)$$

where the loss function is expressed in terms of Kernel functions, namely $K(\boldsymbol{q}_i, \boldsymbol{q}) = \boldsymbol{\phi}^T(\boldsymbol{q}_i)\boldsymbol{\phi}(\boldsymbol{q})$, and $\alpha_i$ is the linear coefficients to be determined through training. Since the loss function of SVM is based on $\mathcal{L}_1$-norm (absolute value), it tends to obtain a more sparse representation (more zero components in $\boldsymbol{\alpha} = [\alpha_1, \alpha_2, \ldots, \alpha_i, \ldots, \alpha_N]^T$) [91]. As a result, SVM generally provides faster predictions with smaller memory use compared with KRR. In contrast, the loss function in KRR is based on the squared error that penalizes the error more heavily during the optimization, thus KRR would generally take less time for the training process. However, both KRR and SVM are limited to small datasets since the evaluation of covariance matrix $K(q_i, q_j)$ scales with the number of data as $\mathcal{O}(N^2)$.

In addition to affecting the performance and accuracy of regression, the choice of loss functions can result in effective feature selections in predicting thermal transport properties. Let us consider another example of high throughput prediction of thermal conductivity directly from material structures and other properties. In this case, each component $q^j$ of the vector $\boldsymbol{q}$ represents a material property such as chemical compositions, Young's modulus, specific heat, bonding length, *etc*. To clearly illustrate how feature selection can be achieved, a linear model is adopted here, namely, the thermal conductivity is expressed as a linear combination of the descriptor features: $k = \beta_0 + \boldsymbol{\beta}^T \boldsymbol{q}$. From physical intuitions, one can expect that significant features affecting thermal conductivity include Young's modulus, atomic masses, and Grüneisen parameters. It is reasonable to expect the corresponding linear coefficient $\beta_j$ should have a large absolute value, if the material property $q^j$ affects thermal conductivity significantly. If a feature $q^j$ such as formation enthalpy is irrelevant to thermal conductivity, the corresponding coefficient $\beta^j$ should be small or even zero. Proper choice of loss functions can automatically set the coefficients of irrelevant coefficients to zero during the training process, which is equivalent to drop these redundant features. A typical example is the loss function of the LASSO (least absolute shrinkage and selection operator) algorithm [92], defined as a constrained optimization problem:

$$L(\beta_0, \boldsymbol{\beta}) = \sum_i (y_i - \beta_0 - \boldsymbol{\beta}^T \boldsymbol{q}_i)^2, \text{ subject to } \|\boldsymbol{\beta}\|_1 = \sum_{k>0} |\beta_k| < t \tag{14}$$

In contrast to the KRR based on $\mathcal{L}_2$ regularization, the regularization here is based on $\mathcal{L}_1$ norm (denoted as $\| \|_1$, which is the sum of absolute values of components), such that most of the components of $\boldsymbol{\beta}$ tend to be set to zero ("sparse"). Such idea of using $\mathcal{L}_1$ norm to find sparse representations of data has been widely applied in efficient signal sampling, processing, and reconstruction, known as compressed sensing [91].

In addition to the algorithms like KRR, SVM, and LASSO that minimize certain loss functions, Gaussian process regression (GPR) [93] is another type of ML algorithm using kernel functions. Instead of minimizing the loss function, GPR is based on Bayes' theorem, which is essentially similar to human's learning behavior from experience. We can understand the process using a simple example of cancer testing. For example, a doctor knows from the literature that the chance of a person getting cancer is one in 5000, which we denote as a "hypothesis" $h$ and $p(h) = 0.02\%$. When the doctor sees a patient, he will perform a cancer test to see if the result is positive (denoted as evidence $D$). From experience, the probability of "true positive" result is $p(D|h) = 99\%$, where $D|h$ denotes positive testing results given that the patient has cancer. On the other hand, the chance of a "false positive" is $p(D|\bar{h}) = 0.005\%$, where $\bar{h}$ denotes not having cancer and $p(\bar{h}) = 99.98\%$. Then $p(D) = p(D|h)p(h) + p(D|\bar{h})p(\bar{h}) = 0.0248\%$. Now, if cancer test turns out to be positive, the doctor will then update the probability of the patient getting cancer using the Bayes' theorem: $p(h|D) = p(D|h) \cdot p(h)/p(D) \approx 80\%$. This is exactly how a doctor learns from the evidence to determine the chances of the patient getting cancer. Similar to this cancer testing example, the Gaussian process solves the regression problem as follows. Given the database $\mathcal{D} = \{(q_i, y_i)|i = 1,2,3,...,N\}$, find the best estimate of the function $y^* = f(q^*)$ evaluated at $q^*$, namely the expectation value of $q^*|\mathcal{D}$. In particular, GPR assumes that all the data points collectively obey the normal distribution:

$$\begin{bmatrix} y \\ y^* \end{bmatrix} \sim \mathcal{N}\left(\mathbf{0}, \begin{bmatrix} K_{QQ} & K_{Q*} \\ K_{Q*}^T & K_{**} \end{bmatrix}\right) \tag{15}$$

where the notation $\mathcal{N}(\mu, \Sigma)$ denotes the normal distribution with the mean of $\mu$ and the covariance of $\Sigma$. In the covariance matrix of Eq. (15), $K_{QQ}$ is a sub-covariance-matrix associated with data points in the dataset $Q = \{q_1, q_2, ...\}$, $K_{Q*}$ is the covariance matrix between the unknown data point $q^*$ and training dataset $Q$, and $K_{**}$ is simply $K(q^*, q^*)$. Through the Bayes' theorem, it can be shown that the conditional probability distribution can be written as:

$$y^*|y \sim \mathcal{N}\left(K_{Q*}^T K_{QQ}^{-1} y, \; K_{**} - K_{Q*}^T K_{QQ}^{-1} K_{Q*}\right) \tag{16}$$

One can easily obtain the best estimate (i.e. expectation value $E[y^*]$) and the variance of the $var[y^*]$:

$$\begin{aligned} E[y^*] &= K_{Q*}^T K_{QQ}^{-1} y \\ var[y^*] &= K_{**} - K_{Q*}^T K_{QQ}^{-1} K_{Q*} \end{aligned} \tag{17}$$

The major advantage of GPR is that the propagation of uncertainties can be rigorously computed. GPR is also very versatile since it does not assume any pre-knowledge of the unknown function nor does it require selection of any loss function for minimizing. The limitation of GPR is its high computational cost since it involves inversion of the covariance matrix $K_{QQ}^{-1}$, which takes $\mathcal{O}(N^3)$ operations with $N$ data points in the database [94]. Such scaling behavior limited the application of GPR for big datasets with millions of data points.

In summary, we discussed several kernel-based supervised learning algorithms. By introducing feature mapping or kernel functions, these algorithms can be highly versatile for nonlinear regression problems such as building interatomic potential or thermal conductivity prediction from material properties. Nonetheless, these algorithms can still be regarded as linear models in higher dimensional kernel space. The training process of these kernel-based algorithms is therefore mathematically simple, which involves minimizing certain loss functions. In general, these simple models perform well for small datasets with relatively low dimensionality of material descriptors, but their scaling behavior with the data size $N$ limits their capability for big datasets compared with some nonlinear regression algorithms, which will be discussed in Section 2.3.2.

### 2.3.2. Nonlinear learning algorithms

The previous section covered some simple linear algorithms of supervised learning. Although these algorithms are versatile for nonlinear regression problems through the kernel functions, the training process usually involves inverting large covariance matrices, and the computational costs increase dramatically with a polynomial scaling $\mathcal{O}(N^2)$ or $\mathcal{O}(N^3)$ with the data size $N$. Nonlinear models, on the other hand, avoid such numerical problems. The training is thus much more scalable when dealing with big datasets with millions of data points. For example, the training complexity of a neural network using stochastic gradient descent scales as $\mathcal{O}(KN)$, with $N$ representing the data size and $K$ the number of adjustable parameters.[71] This section introduces two most commonly used nonlinear supervised learning algorithms: artificial neural networks (ANN) and decision tree-based methods, with the former mimicking the biological network of neurons and the latter similar to the human's decision process. These two nonlinear models are both versatile for high dimensional, large size of input data, but with a tendency to overfitting. Variations of ANN and decision tree-based methods such as convolution neural networks and random forests have been developed to overcome the limitations of their pristine models.

Figure 6a shows a simple architecture of a feed-forward ANN. Such neural network starts with an input layer of neurons that takes the input variables $x_i$, followed by several "hidden layers" and finally ends with an output layer. Each layer contains a certain number of neurons, and each layer takes the outputs of the previous layer as input and its outputs are passed on to the next layer as inputs. Such data flow is visualized as connections between the neurons.

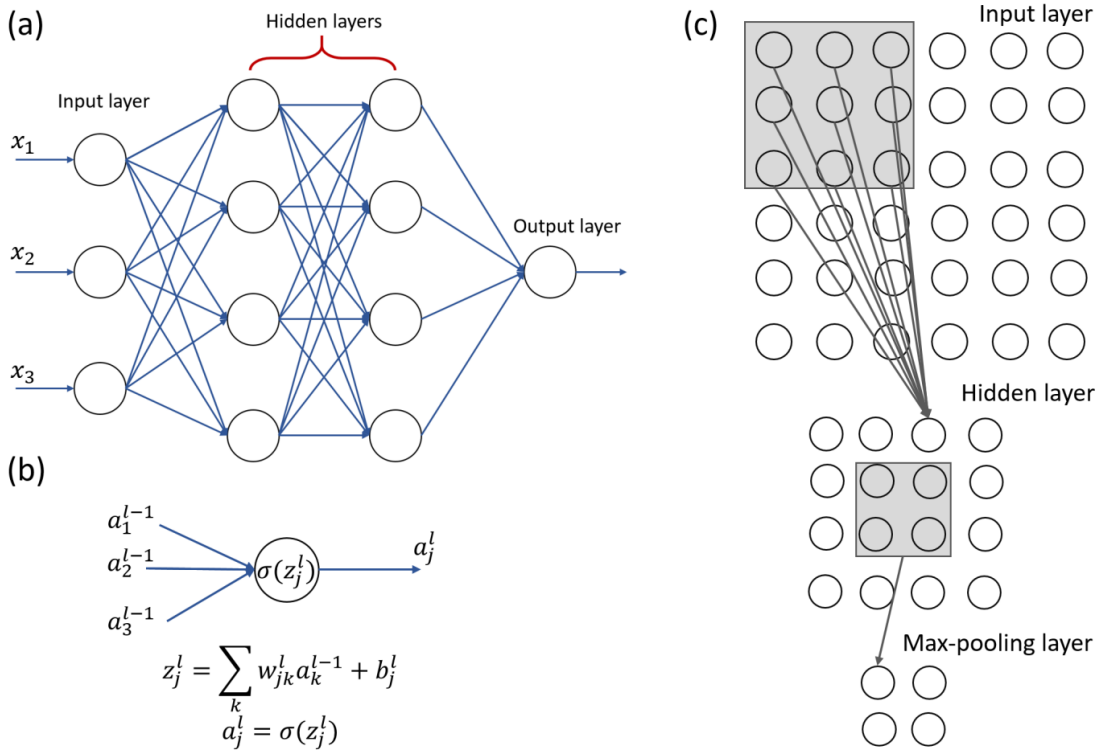

Figure 6. (a) Topology of a three-layer feed-forward neural network. Each circle indicates a neuron and the arrows indicate the data flow between the neurons. (b) Schematic of a single neuron. (c) The topology of the first three layers of a convolution neural network (CNN). CNN involves a set of neurons sharing the same weights and biases with the neuron in the next hidden layer, and max-pooling layers that only take the maximum activation of a certain set of neurons in the last layer as the input.

To illustrate how a neural network works, let's consider a neuron that is $j$-th neuron located $l$-th layer of the neural network, and its mathematical structure is shown in Figure 6b. Such neuron is associated with an activation function $a_j^l = \sigma(z_j^l)$, which is to quantify the degree that the neuron is activated ($a_j^l$) when it receives an excitation $z_j^l$. One of the most well-known choice of the activation function is the sigmoid function $\sigma(z) = (1 + e^{-z})^{-1}$. The excitation $z_j^l$ is a linear combination of all the activation of the neurons in the last layer:

$$z_j^l = \sum_k w_{jk}^l a_j^{l-1} + b_j^l \tag{18}$$

where $w_{jk}^l$ is the weight associated to the connection between $k$-th neuron in the $(l-1)$-th layer and the $j$-th neuron in the $l$-th layer, and $b_j^l$ is called "bias" associated with the neuron which determines the baseline of activation level. The weights determine the strengths of the correlation between the neurons. Different distributions of the weights in the neural network determine the pattern of how neurons become activated by a given input, and how such excitation propagates through the network for obtaining the final result $y$. The goal of the training process is therefore determining the optimal numerical values of the weights and the biases, such that the neural networks would output results consistently with the training datasets. Numerically, training of the neural network is to find the optimal values of the weights and biases that minimizes the error between the output layer and the reference outputs $y_j$. The loss function for the minimization can then be defined as:

$$L = \frac{1}{2n} \sum_{x \in X} \sum_j \|a_j^L - y_j\|^2 \tag{19}$$

where $a_j^L$ is the activation of $j$-th neuron of the output layer $L$, $\boldsymbol{X}$ the vector containing input variables $x$ in the training dataset, and $n$ the number of training data entries. The training can be achieved through a method called back-propagation which is based on gradient descend minimization [71]. However, one significant challenge for training deep neural networks with more than 5 layers is that the gradient of the weights and biases vanishes as the number of connections between hidden layers increases.[71] This would significantly slow down the training process. The problem of the vanished gradient in deep neural networks can be mitigated by choosing different forms of activation functions such as rectified linear unit, sigmoid linear unit, *etc.* [95, 96], or by using different loss functions with larger gradient scaling behavior with the number of hidden layers. One most used loss function other than the quadratic form in Eq. (19) is the "cross-entropy" [71]:

$$L = -\frac{1}{n}\sum_{x \in X}\sum_{j}\left[y_j \ln a_j^L + (1 - y_j)\ln(1 - a_j^L)\right] \quad (20)$$

Intuitively, an increased number of hidden layers would increase the "flexibility" for the neural network model when dealing with more complicated regression problems, while it also results in higher costs of training and the tendency of overfitting when the size of the training dataset is limited.

Convolution neural networks (CNNs) [71, 97] with a different architecture as shown in Figure 6c can circumvent the problem of the vanished gradient and suppress overfitting compared with fully connected feed-forward neural networks. Instead of using fully connected neurons, some arrays of the neurons in the input layer share the same weights and biases with the first hidden layer. Following the first hidden layer, there is a max-pooling layer, which only takes the maximum activation as the input to the next layer. This could reduce the number of degrees of freedom for minimizing the loss function during the training process, which suppresses the decay of the gradient across the number of layers. The other merit of these pooling layers is that they can be effectively regarded as an intrinsic feature selection process: only the most important neuron connections yielding maximum activations survive through the max pooling. Such built-in feature selection in CNN makes it robust even when the material descriptor is not complete, since CNN intrinsically improves material representations in the convolution and max-pooling layers [83].

Neural networks are generally flexible for regression problems. According to the universal approximation theorem [71, 98], for any arbitrary piecewise smooth function $f$, it is guaranteed to exist a neural network that can achieve arbitrary desired precision to compute $f(x)$ for every possible input $x$. Neural networks are thus most suitable to model high-dimensional nonlinear functions with a fast speed of prediction once trained. However, neural networks are generally black-boxes in contrast to simple linear kernel-based algorithms such as GPR: it is usually challenging to interpret the functionality of certain neurons in correlating the output and the input variables.

Another family of commonly used nonlinear regression models involves decision trees [99, 100] as shown in Figure 7. In contrast to the neural network that is a "black-box" model, a decision tree is a "white-box" model how the final output is reached through the decision process is quite visible. Figure 7a shows an example. If a decision tree is used for screening thermodynamically stable crystals, the decision tree will ask a series of logical questions, such as whether the formation enthalpy is positive and whether all phonon frequencies are positive. Each question aims at separating the input data into subgroups until the tree can decide the crystal is stable, metastable, or unstable (Figure 7c). Such classification can be visualized as a tree-shaped flow chart, with internal nodes represent a logical condition aiming at dividing data into sub-classes, and leaf nodes that no further split of data is performed and predictions are made. In addition to performing classification, decision trees can also be used for regression problems, which are referred to as regression trees. The regression tree fits the unknown function by deciding which interval the input variable belongs to, as shown in Figure 7b. For the special tree structure as shown in Figure 7c, the input variables are divided into five intervals and the boundary of the intervals are called 'decision boundaries'. The decision tree regression seeks the best decision boundaries that minimize the following loss function:

$$L = \sum_k \sum_{i \in C_k} (y_i - m_k)^2 \tag{21}$$

where $C_k$ denotes the subclass represented by the leaf node $k$, $m_k = \frac{1}{N_k} \sum_{i \in C_k} y_i$ is the prediction made by the $k$-th leaf node, which is essentially the mean value of the data belong to this subclass. Unlike regression algorithms discussed before, decision trees only result in piecewise constant functions as shown in Figure 7b. One important disadvantage of the decision tree is that it tends to overfit if the maximum number of nodes is not specified. To solve this problem, the random forest approach has been developed [101].

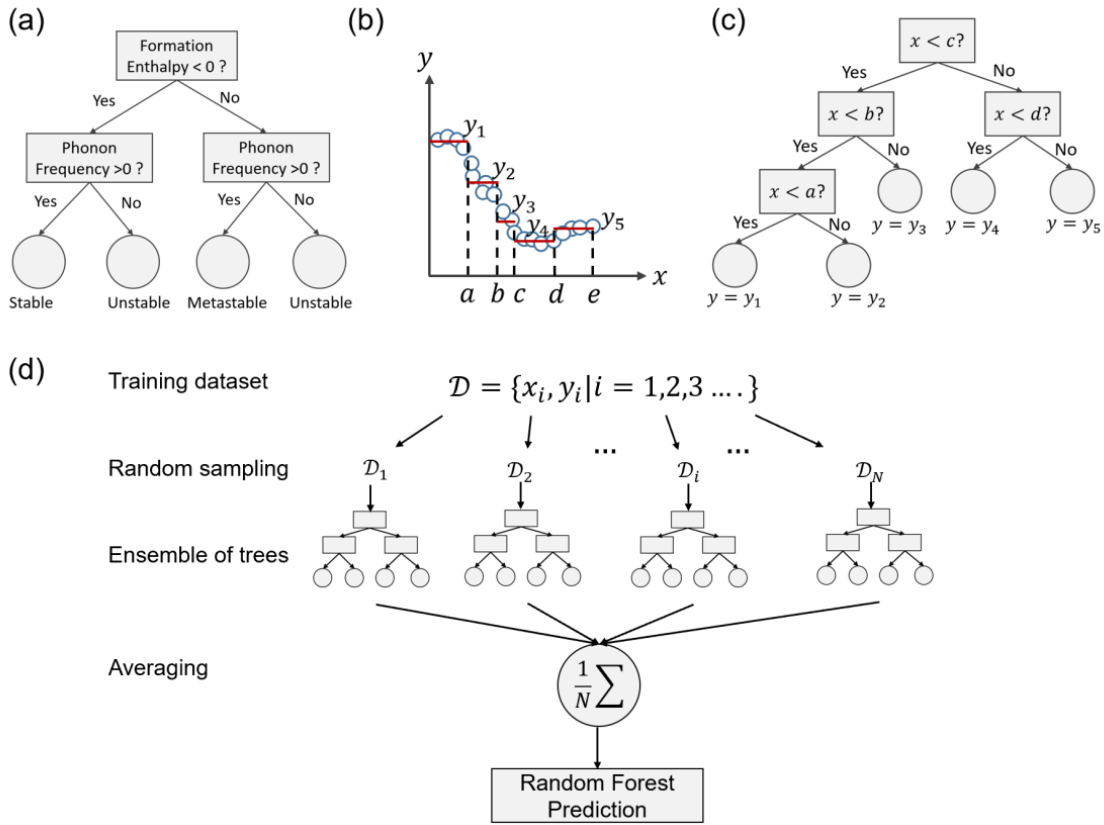

Figure 7. Decision trees and random forest. (a) a classification tree for determining the stability of materials. (b) A regression problem that is fitted by (c) a regression tree. (d) Schematic of a random forest. From the training dataset, random sampling is performed to obtain training batches $\mathcal{D}_1, \mathcal{D}_2, ..., \mathcal{D}_N$ with each batch used to train a decision tree. Then the ensemble of trees is averaged to make a prediction. Note that, each data point can be sampled multiple times.

Instead of relying on just one decision tree, a random forest involves an ensemble of trees, as shown in Figure 7d. Each tree is trained based on a subset $\mathcal{D}_i$ of data, which contains randomly sampled data from the database. Note that repeated sampling of the same data entry is allowed when constructing the training subsets, which is known as the Bootstrap method [102]. Each tree can be cross-validated using the datasets used for the training of other trees. When making predictions, the final output is averaged among all the decision trees, which also suppressed the stochastic error through the ensemble average. Therefore, the random forest is commonly used for predicting material properties such as thermal conductivity with limited data availability, which will be discussed in the following sections. The random forest can also be used for feature selection. The idea is to remove a certain feature $q^k$ from the original descriptor $q$ to form a new database, and see how much the feature removal process would affect the fitting error of the regression using the random forest trained with the new database. For example, if the squared error for thermal conductivity prediction increased a lot by removing atomic masses from the material descriptor, it means atomic mass will be an important feature for approximating thermal conductivity. A similar process can be performed for all the features in the descriptor vector to rank the importance of all features in this way. This strategy of feature selection is referred to as recursive feature elimination [103], which can also be applied along with simple linear methods like SVM. [104]

Finally, it is important to note that the selection of ML algorithms is case-sensitive. Clearly, the size of the dataset plays an important role in the aspects of overfitting and underfitting. Therefore more complex and flexible models like neural networks are preferred with the increased size of data, but they are also more likely to have overfitting problems due to the high variance nature. It is also recommended to start with simple models like KRR as a benchmark of performance before spending much time training complex models with many adjustable parameters such as deep neural networks.

### 3. Applications of machine learning in thermal transport properties

In this section, recent progress in the applications of ML algorithms in modeling thermal transport properties is reviewed, especially those addressing key challenges in understanding thermal transport physics, high throughput screening, and the designing of functional materials with target thermal transport properties.

### 3.1. Machine learning potentials for thermal transport simulations

As discussed in the Introduction, *ab initio* phonon simulation based on BTE and the perturbation theory faces challenges in modeling thermal transport in materials at high temperatures, in materials with defects such as alloys, and in amorphous materials. MD, on the other hand, is a convenient choice since it does not rely on the phonon gas model but directly simulates the movement of atoms. Unfortunately, the timescale required for modeling thermal transport should be at least longer than the lifetime of the vibrational modes, usually on the order of hundreds of picoseconds, which is beyond the reach of DFT-based *ab initio* molecular dynamics (AIMD). Machine learning potentials (MLPs) provides a promising solution to bridge the gap between DFT and MD simulations. MLP is a type of interatomic force field that employs ML regression algorithms trained using *ab initio* data to fit the correlation between atomic configurations and their energies. Unlike empirical potentials that are commonly used in conventional MD simulations, MLP does not need the assignment of functional forms. Instead, MLP adaptively fits the PES through "learning" from the training dataset. Such advantages of MLP could significantly improve the accuracy in computing energies and forces compared with the empirical potentials.

The birth of MLP can be dated back to the work by Blank *et al.* in 1995 [105]. They used a three-layer feed-forward neural network to model CO molecule adsorption on Ni (111) surfaces. The descriptor used in this work was very simple, mainly the center-of-mass coordinates of the CO molecule, and the angle between the C-O bond and the Ni (111) surface. This pioneering work demonstrated that ML methods could be used to make accurate predictions of the PES of a simple molecular system with only a few degrees of freedom. For condensed matter systems, however, the PES is a high-dimensional function depending on the position of all the atoms. One of the major advancements in developing MLP is the work by Behler and Parrinello in 2007 [81]. The authors assumed that the total energy of simulation system $E$ can be decomposed to atomic energies ($\varepsilon_i$): $E = \sum_i \varepsilon_i(\boldsymbol{q}_i)$, and the atomic energy $\varepsilon_i$ of a certain atom $i$ is a local property determined only by the configuration of its neighboring atoms fingerprinted by $\boldsymbol{q}_i$. This assumption allows the application of MLPs to condensed matter systems with a large number of atoms. Specifically, Behler and Parrinello have proposed symmetry-invariant pairwise and three-body potential functions to construct descriptor vectors. In 2012, Bartók *et al.* provided a rigorous theoretical analysis on how to construct a complete set of a descriptor that fully incorporates many-body interactions[106]. Although SOAP descriptor proposed by Bartók *et al*. [107, 108] was implemented with the Gaussian approximation potential (GAP) package, it can also be used as descriptors with other machine learning regression algorithms [109]. Many MLPs have been developed by combining descriptors of atomic configurations and supervised learning regression since then. In addition, MLP models based on other ML algorithms including GPR [94], compressed sensing[110], SVM[111] have been developed. In general, these MLPs can achieve high accuracy in predicting atomic energies (0.1 meV/atom) and interatomic forces (0.01 eV/Å) [107]. These MLPs have been successfully used to model the structures and mechanical properties of simple crystals such as Si [81, 107, 112], GaN [107], and graphene [113], as well as complex atomic structures and processes, such as the amorphous carbon [114], lithium-ion transport in electrode materials [115, 116], and phase-change material GeTe [117]. The advancement in developing MLP has been summarized in the review articles by Behler [118] in 2016 and then by Mueller [119] in 2020. However, studying phonon properties and thermal transport using MLP is still at its infancy.

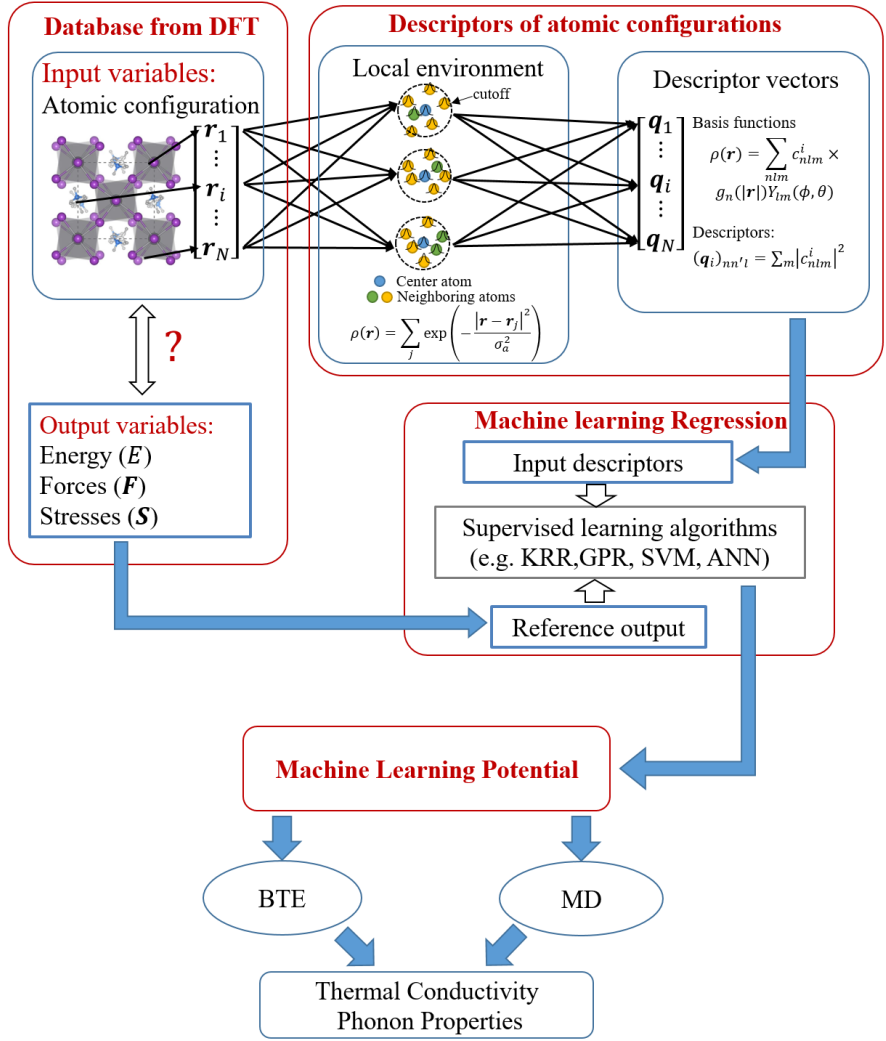

Figure 8. The workflow for applying supervised learning to develop machine learning potentials (MLPs) for predicting thermal conductivity and phonon properties.

Figure 8 shows the workflow for developing MLP to study thermal transport and phonon properties. Developing an MLP is essentially approximating the function $f(\mathcal{X})$, whose variable is simply atomic configuration $\mathcal{X}$ and the output variables are energies $E$ or interatomic forces $\boldsymbol{F}$. A critical step shown in Figure 8 is to find appropriate descriptor vectors in the mathematical form to uniquely fingerprint the atomic configurations, as discussed in Section 2.2. After converting the atomic coordinates to descriptor vectors $q$, these descriptors are used as the reference input variables for training the MLP. Reference output variables including the energy and interatomic forces are computed using DFT, which is also included in the database. The goal of the database is to sample the PES in the space of atomic configurations, and one way to achieve such sampling is to perform *ab initio* molecular dynamics (AIMD) and record the trajectory of atoms and the potential energy at each AIMD snapshot. Alternatively, PES can be sampled by imposing perturbations to the crystal and computing the corresponding energy and forces. Examples of perturbations are randomly displacing atoms away from equilibrium or applying strains to the crystals. After obtaining the training database, one can select the machine learning algorithms to build an MLP. One can then either directly run MD simulations [120-125] or use MLP to calculate force constants and solve BTE [126-128] to calculate thermal conductivity and perform further analysis of phonon properties.

### 3.1.1. Transport and vibrational properties at high temperatures

Understanding the thermal properties of materials is important for a lot of high-temperature applications, such as thermal barrier coatings, nuclear applications, and high-temperature thermal insulation materials. Prediction of the thermal properties requires accurate microscopic descriptions of the vibrational spectra of solids. However, it has been a longstanding challenge in condensed matter physics to precisely extract phonon properties at high temperatures, especially for materials with phase changes whose high-temperature atomic structure differs from the ground state structure. When analyzing the vibrational spectra and phonon dispersions of these high-temperature phases, conventional theoretical frameworks of the finite displacement method [129] or density functional perturbation theory [130] usually predict that these material phases are dynamically unstable. On the other hand, MD is a promising tool for studying high-temperature properties in a nonperturbative manner, since it directly models the trajectories of atoms at high temperatures. However, the limitation of MD lies in the accuracy of empirical potentials. MLP, being able to predict interatomic forces with accuracy close to that of DFT, solved the limitations of fidelity and accuracy of empirical potentials, therefore provide a promising tool for study thermal properties at high temperatures. The research by Qian and Yang provides a proof-of-concept study on utilizing MLP to analyze high-temperature vibrational properties of Zr [131]. Based on the GAP framework developed by Bartók [107], Qian and Yang trained an MLP model [131] based on SOAP descriptors and GPR algorithm for elemental Zr crystal through AIMD trajectories. They then investigated the phonon dispersion and anharmonic effects in Zr crystal at high temperatures [132]. As shown in Figure 9a, conventional lattice dynamics based on the finite displacement method [129] predicted soft phonon modes with imaginary frequencies in the body-centered cubic phase (bcc) of Zr. Such soft phonon modes suggest that the bcc phase is dynamically unstable. However, the bcc structure is observed experimentally as a stable phase of Zr above 1188 K. In 1955, Hooton realized that atoms vibrate in an effective potential due to their nonstationary neighbors, and the PES is stochastically sampled around the most probable position which is not necessarily a local minimum [133]. Such thermal vibrations would renormalize the soft phonon modes to phonon modes with real frequency at high temperatures due to the existence of anharmonicity. As shown in Figure 9b, the PES shows a clear double-well shape for the soft transverse acoustic (TA) mode at the N point. The bcc structure is indeed a local maximum of the PES. As a result, a positive-indefinite dynamical matrix would be obtained if one tries to use a harmonic Hamiltonian to approximate such PES. At high temperatures, the amplitude of oscillation is larger so that the oscillator would hop between the two local minima as shown in Figure 9b, and the bcc structure of Zr is a dynamical average between the two local minimum. Along this line, Qian and Yang performed MD simulation at high temperatures and have successfully obtained the renormalization of soft phonon modes in bcc Zr as shown in Figure 9c [132], with excellent agreement with inelastic neutron scattering results as shown in Figure 9d.

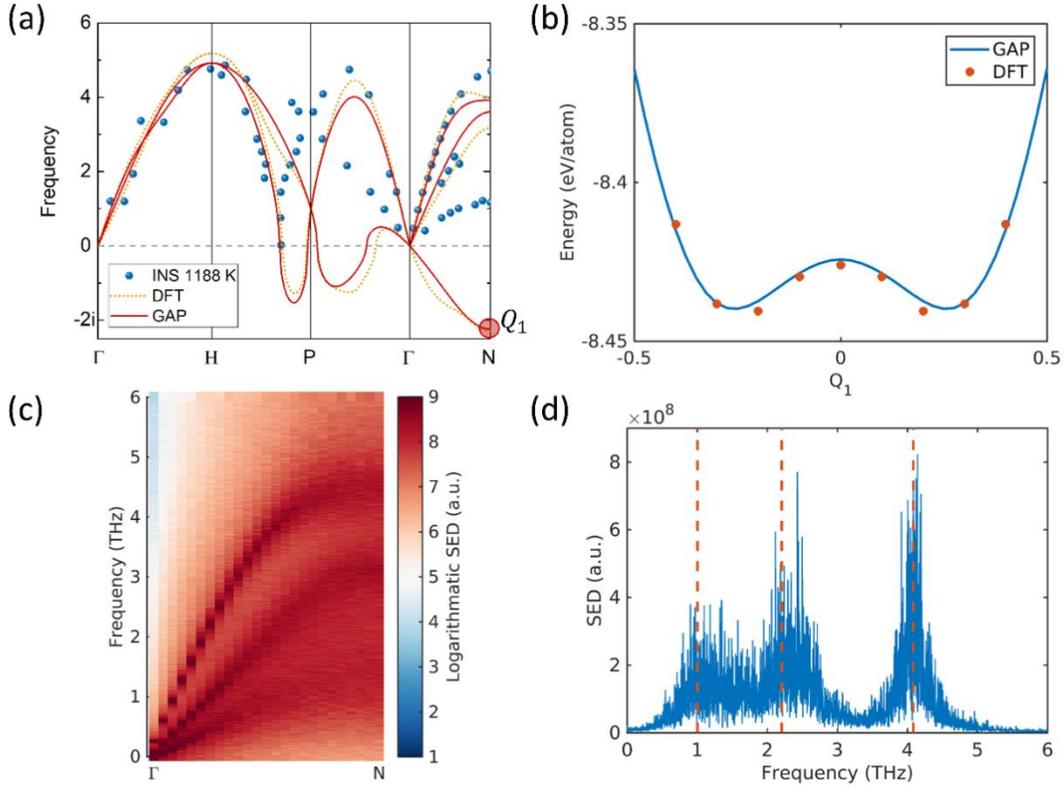

Figure 9. Renormalization of soft phonon modes in bcc-Zr modeled using Gaussian approximation potential (GAP). (a) Phonon dispersions calculated by the finite displacement method. (b) Potential energy surface for the soft transverse acoustic (TA) mode at N point on the boundary of Brillouin zone. $Q_1$ denotes the dimensionless normal coordinate. (c) Phonon dispersion obtained using the spectral energy density (SED) method. (d) SED at a fixed wavevector $\bm{q}$ =(0.3,0,0), along the $\Gamma$-N direction. Reproduced from ref. [132], Copyright 2018 by American Physical Society.

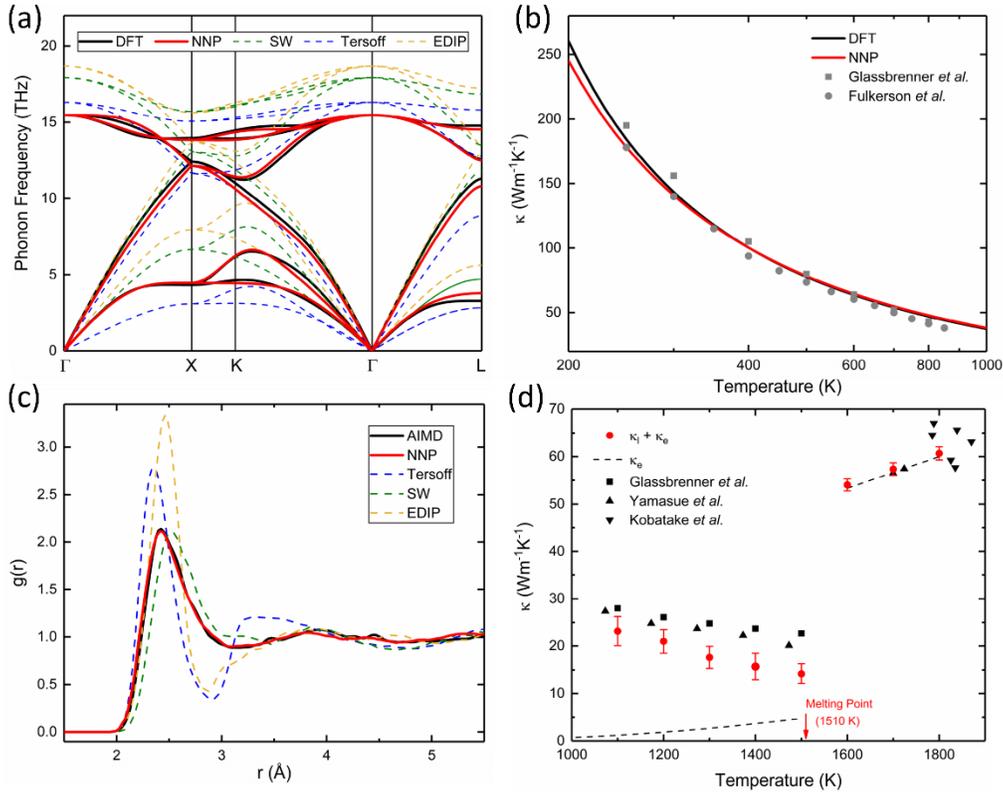

Figure 10. (a) Phonon dispersion of crystalline silicon using a neural network potential (NNP), compared with DFT simulations and empirical potentials such as Stillinger-Weber (SW) [134], Tersoff [135], and the environment-dependent interatomic potential (EDIP) [136]. (b) Thermal conductivity of crystalline silicon up to 1000 K. (c) Radial distribution function of liquid silicon. (d) Total thermal conductivity of crystalline and liquid silicon near the melting point including lattice contribution $k_l$ and electronic contributon $k_e$. Experimental values of thermal conductivity are by Glasbrenner *et al.* [137], Yamasue *et al.*[138] and Kobatake *et al*. [139]  Figure reproduced from ref. [140], Copyright 2019 by Elsevier.

In addition to solids, MLP has been recently used to model liquids as well. Recently, Li *et al.* developed a unified deep neural network potential (NNP) to study thermal transport in Si below and above the melting point [140]. The framework based on lattice dynamics and the BTE which describes the phonon gas picture would break down for studying thermal conductivity of systems with solid-liquid phase change. Their work demonstrated the flexibility of the deep neural network for fitting PES across a wide range of atomic configurations in distinctly different phases. The deep neural network potential is trained by randomly displaced Si structure at 0 K as well as AIMD trajectories ranging from 50 K up to 3000 K. Figure 10a-b showed excellent agreement with DFT simulation for both phonon dispersion and the temperature-dependent thermal conductivity of crystalline Si, from 200 K up to 1000 K. In addition to thermal conductivity, the radial distribution function of liquid Si is also reproduced as shown in Figure 10c. Figure 10d summarizes the thermal conductivity of silicon near the melting point. Using EMD, the vibrational contribution to thermal conductivity $k_l$ is estimated to be only 0.7 W/mK for the liquid phase, and the majority of thermal conductivity for liquid silicon comes from electronic contributions.

While these preliminary researches focused on elemental materials, MLP has shown its predictive power of modeling thermal properties, especially for materials with phase transitions at high temperatures, and MLP will be a promising tool for screening and prediction of thermodynamic stability and thermal conductivity of advanced materials for high-temperature applications, such as molten salts, high entropy oxides, and alloys.

### 3.1.2. Thermal transport in crystals with defects

Conventional methods for modeling phonon-defect interaction are usually based on virtual crystal approximation (VCA). VCA is essentially a mean-field treatment by viewing the alloy as an effective "virtual crystal" with mean values of mass and bonding force constants, and the phonon-defect scatterings are attributed to the mass disorder [141, 142]. Such assumptions are questionable when there exists significant lattice distortion, which causes local stress field and inhomogeneity of bonding stiffness [38]. As a result, VCA could overestimate the thermal conductivity of alloys [143, 144]. Until recently, phonon BTE and atomistic Green's function have been integrated together to capture the local distortion and relaxation of atomic structures near the defects [26, 39]. However, this BTE-based method would still be questionable when the concentration of random defects is high, as the phonon gas model could break down in these strongly disordered materials [143, 145]. The recent development of MLPs provides an opportunity for direct modeling of defects and their effect on atomic vibrations and thermal transport without any mean-field assumptions because ML regression algorithms can fit the PES over a vast landscape of atomic configurations due to the flexibility. Gu and Zhao developed an MLP to study thermal transport in alloyed two-dimensional material $MoS_{2(1-x)}Se_{2x}$ [146]. The method they used is referred to as spectral neighbor analysis potential (SNAP) [147], which is essentially a linear regression model in the space of descriptors. Their SNAP model is trained using DFT-derived energies, forces, and stresses of the strained and randomly displaced simulation cells of $MoS_{2(1-x)}Se_{2x}$ with different compositions. As shown in Figure 11a-b, the SNAP potential can properly model phonon properties of $MoS_2$ and $MoSe_2$ compounds even when they form superlattices. Since atomic details are naturally incorporated in MLP, they showed that force-field disorder contributes significantly to thermal resistance in the alloy system, as shown in Figure 11c. Figure 11d shows the participation ratio of vibrational modes. A lower participation ratio is found in $MoS_{2(1-x)}Se_{2x}$ alloy with x=0.5 when both mass disorder and force-field disorder are taken into account, which indicates stronger localization of the vibration modes.

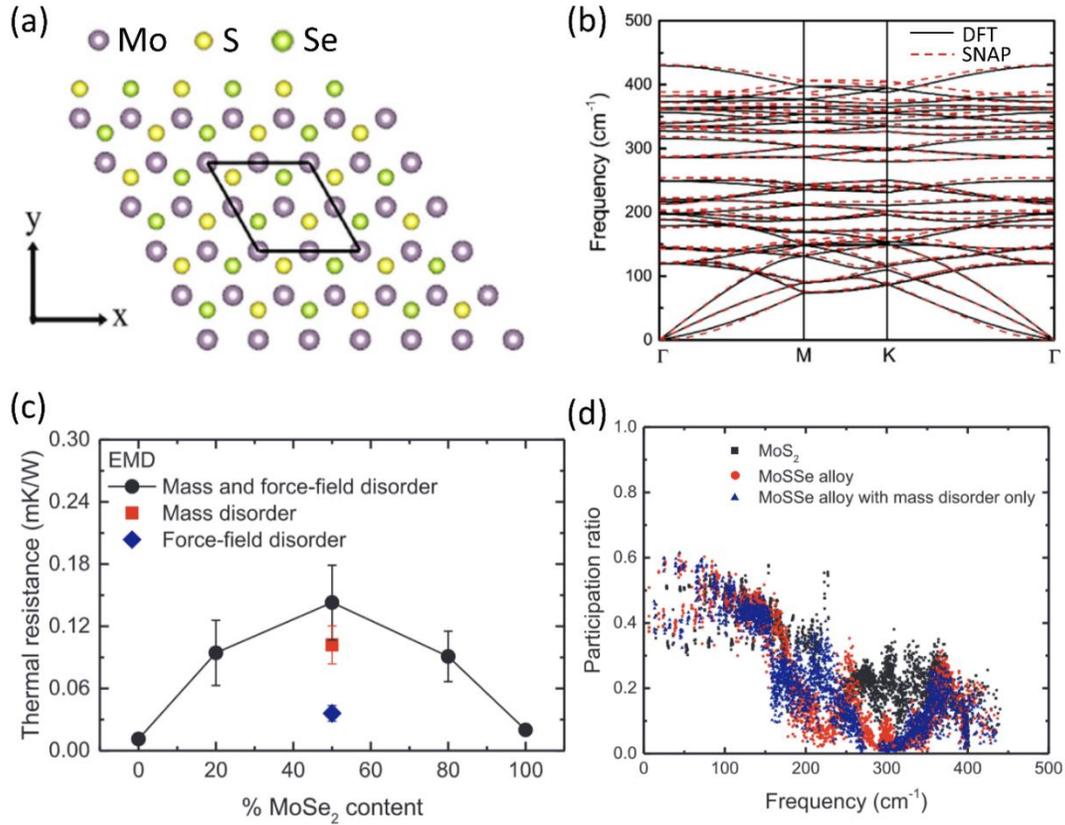

Figure 11. MLP for modeling thermal transport in alloys MoSSe alloy system. (a) A supercell of MoSSe superlattice. (b) Phonon dispersion of MoSSe superlattice. (c) The dependence of thermal resistances of $MoS_{2(1-x)}Se_{2x}$ alloy on force-field disorder and mass disorder. (d) Participation ratio of MoSSe alloy compared with $MoS_2$. This figure is reproduced from ref. [146]. Copyright 2019 by Elsevier.

Another demonstration of using MLP for modeling thermal transport in the disordered system is by Babaei *et al*.[148] They developed a single GAP model for studying thermal transport in both perfect crystalline Si and Si with atomic vacancies. Instead of directly performing MD simulations, the authors showed that thermal conductivity can also be calculated using BTE with interatomic force constants derived from the trained GAP model and vacancy scattering rates derived from atomistic Green's function [39, 149].

### 3.1.3. Thermal conductivity of amorphous materials

Compared with alloys that have substitution disorders, the topology of atomic bonds in amorphous materials completely breaks down the translational periodicity. Therefore, describing thermal transport in amorphous materials requires a completely different theoretical framework beyond the phonon gas model. As coined by Allen-Feldman (AF) model based on the quantum linear response theory, vibrational modes in amorphous materials can no longer be simply regarded as propagating lattice waves (phonons). Instead, new vibrational features appear for systems without periodicity as shown in Figure 12a-b. Not all extended modes are propagating lattice waves (propagons), while some of them become diffusons without well-defined group velocities and wavevectors. These modes contribute to thermal conductivity through random hopping or scattering with propagons. At the high-frequency end of the spectra, there are localized vibration modes called locons. Due to such complex characteristics of vibrational modes, BTE is no longer applicable to predict the thermal conductivity of amorphous solids, and modeling the thermal conductivity in amorphous materials has long been relying on MD simulations with empirical potentials. However, empirical potentials usually failed to correctly predict the atomic structures of amorphous materials, such as pair distribution functions [150] and the percentage of different bonding types [114]. Because MLPs can smartly fit the PES even with highly diverse and complex local atomic configurations [151], MLPs provide a promising way to model the thermal transport of amorphous materials with high fidelity.

The first demonstration of applying MLP to study thermal transport in amorphous materials with strong disorders was by Campi *et al.* [152], where thermal conductivity of the phase change material GeTe is studied using both equilibrium molecular dynamics (EMD) and nonequilibrium molecular dynamics (NEMD) based on neural network potential (NNP). As shown in Figure 12c, the bonding of Ge atoms in GeTe can be generally classified as tetragonal coordination and defective octahedron coordination [153]. The NNP trained by Campi *et al.* [152] successfully captured such complicated geometry of disordered bonding network with an excellent agreement of vibrational density of states with DFT shown in Figure 12d, demonstrating that MLP can smartly fit the PES according to the local chemical environments. Figure 12e shows that a major contribution to thermal conductivity comes from diffusons.

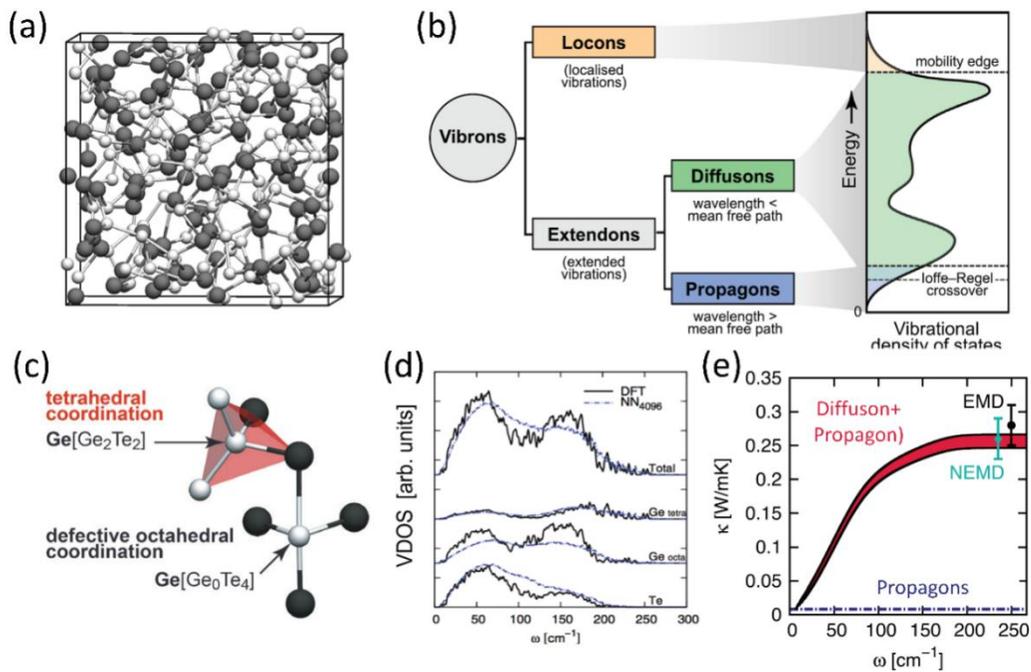

Figure 12. (a) Structure of amorphous GeTe. (b) Vibrational modes in amorphous materials according to Allen-Feldman theory. (c) Coordination types of Ge in amorphous GeTe. (d) Vibrational density of states (VODS) obtained by DFT and neural network with a simulation cell of 4096 atoms (NN4096). (e) Thermal conductivity of amorphous GeTe obtained using EMD and NEMD. (a) and (c) are taken from ref. [153] Copyright 2014 by Wiley, (b-e) are reproduced from ref. [151], Copyright 2018 by Taylor & Francis Group.

Recently, Qian *et al.* [150] developed a GAP model for predicting the thermal conductivity of amorphous silicon. Their model is trained by generating random displacements that stochastically sample the PES. The PES can be effectively sampled with only 150 DFT calculations compared with AIMD with a few thousand steps. Their model reproduced interatomic forces and captured the structural properties of a-Si where a much better agreement with DFT was found compared with empirical potential as shown in Figure 13a-b. Excellent agreement in thermal conductivity has also been achieved between the prediction by MLP and literature values as shown in Figure 13c, demonstrating a promising application of MLP in studying thermal transport in amorphous systems with no empirical fitting parameters.

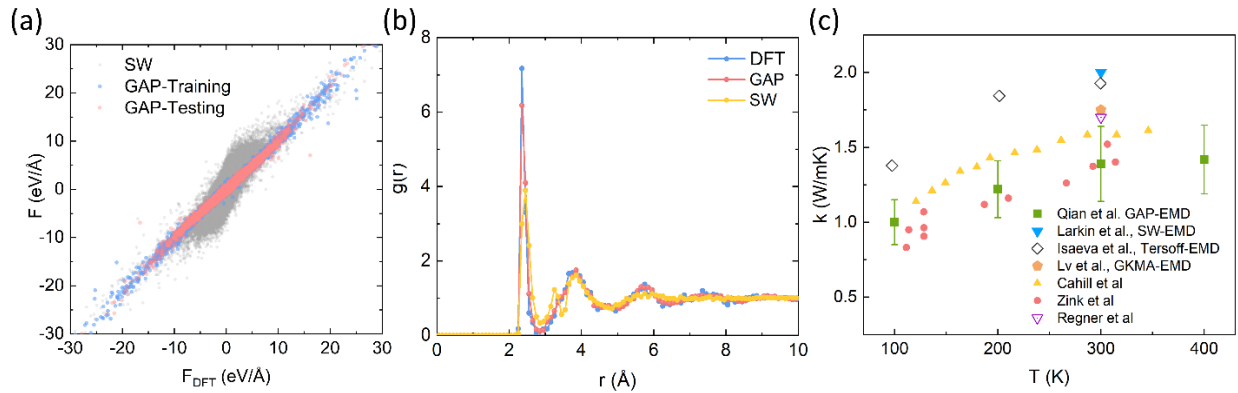

Figure 13. GAP for modeling thermal transport in amorphous silicon. (a) Accuracy of reproducing the interatomic forces using GAP and empirical Stillinger-Weber (SW) potential. (b) Radial distribution function obtained using DFT, GAP, and SW. (c) Thermal conductivity predicted using GAP and EMD, in comparison with experiments by Glassbrenner *et al.*[137], EMD simulation using SW potential by Volz *et al.*[154], and BTE simulation using SW potential by Babaei *et al.* [148]. (d)Thermal conductivity of amorphous silicon calculated using EMD with the GAP model developed in this work, compared with EMD results by Larkin *et al.* [155], Isaeva *et al.* [42], and Lv *et al.* [125], and experimental measurements by Zink *et al.* [156]., Regner *et al.* [157] , and Cahill *et al.* [158]. This figure is reproduced from ref. [150], Copyright 2019 by Elsevier.

## 3.2. Machine learning-driven high-throughput screening

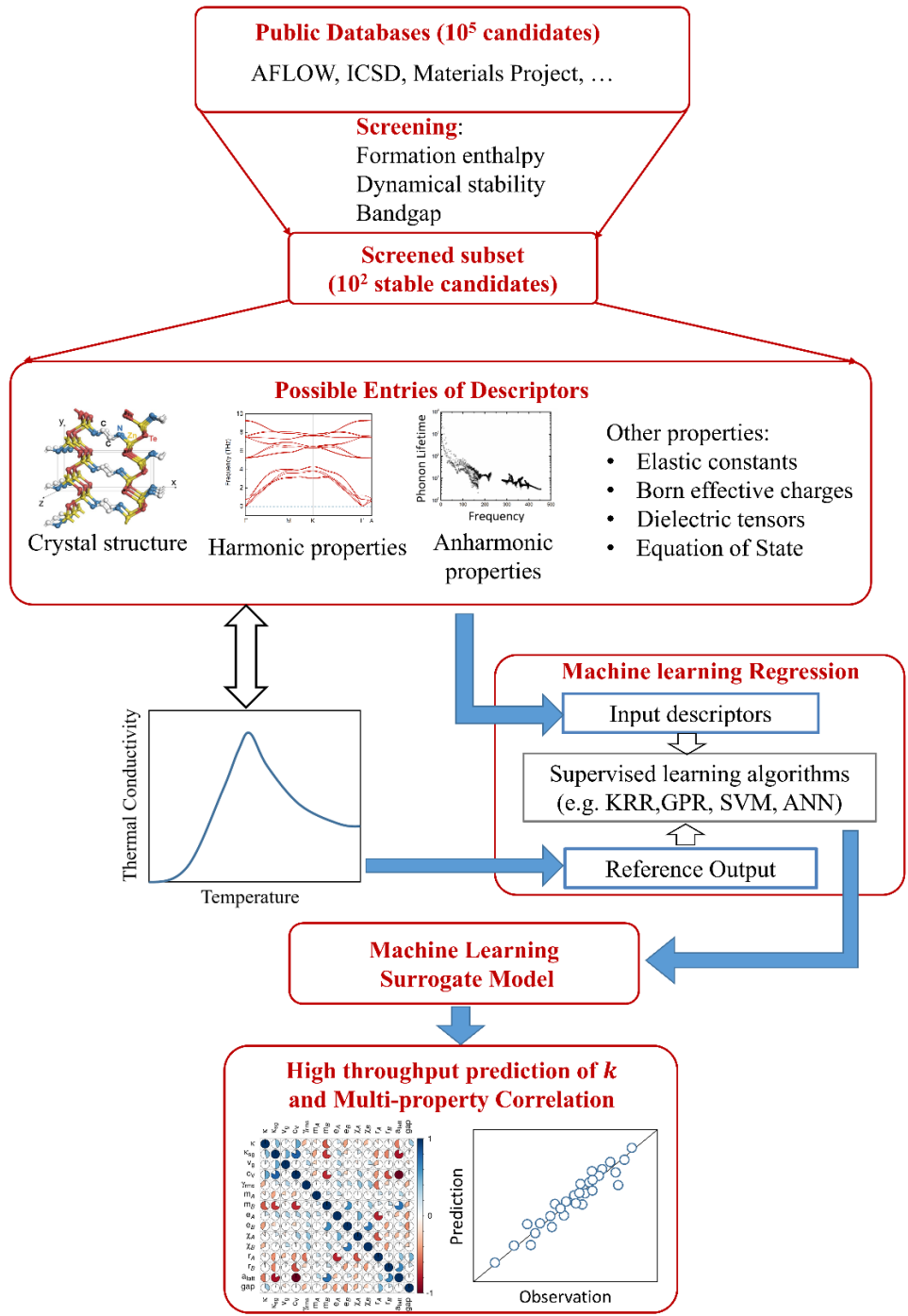

Figure 14. Machine learning as a high throughput prediction and screening tool for thermal conductivity of functional materials, showing thermal barrier coating (TBC) as an example. The correlation graph inset is adapted from ref. [159], Copyright by American Physical Society 2016.

This section focuses on how ML is applied to efficiently select material with superb thermal conductivity from a large pool of materials rather than just analyzing fundamental transport physics. ML-driven high-throughput screening aims at generalizing the correlation between thermal conductivity and other possible material properties from the training dataset and to efficiently predict the thermal conductivity of candidate materials. Along this line, Figure 14 shows the general strategy of combining machine learning and high throughput screening (HTS) of materials with target thermal transport properties. HTS is started with a prescreening process through a series of requirements and criteria to rule out unpromising candidate materials. For example, screening criteria of stable thermoelectric materials usually include negative formation enthalpy for thermodynamic stability, free of soft phonon modes for dynamical stability, and nonzero bandgaps. Such prescreening could significantly reduce the range of candidate materials, usually by a few orders of magnitude. The remaining pool of materials, however, might still be too many for *ab initio* prediction of thermal conductivity due to the computational cost. Machine learning algorithms can therefore be used here as a surrogate model for first-principles phonon calculations to perform high-throughput prediction and screening of thermal properties. However, it is important to note that, such high throughput prediction and screening only seek to capture the general trend of thermal conductivity with the descriptors. In this case, deviations of predicted thermal conductivity values from reference values can be on the order of 300% [54]. Similar to building MLP, training machine learning models for HTS purposes also involves a selection of features and machine learning algorithms.

As discussed in Section 2.1, the major challenge for HTS of thermal transport properties is the limited size of data related to phonon properties, especially anharmonic properties. One way to tackle the challenge of lacking anharmonic properties is presented by Carrete *et al.* through transferred estimation of anharmonic force constants [55]. The authors assumed that anharmonic force constants are more or less transferable across similar crystalline structures, despite the different elemental compositions. For example, they estimated the anharmonic force constant of different half-Heusler (HH) compounds using the anharmonic force constants of $Mg_2Si$, and only perform *ab initio* simulations of harmonic force constants. In this manner, one can significantly reduce the amount of computational cost and afford to calculate anharmonic phonon properties and thermal conductivity for up to 75 HH compounds [55]. They showed that accurate estimation of thermal conductivity can be achieved with up to 93% of the spearman correlation coefficient between the transferred estimation and the results using anharmonic force constants from first-principles calculations.

The selection of descriptors for high-throughput screening should be carried out carefully depending on different types of material properties. For example, Legrain *et al*. showed that thermodynamic properties such as vibrational free energy and entropy can be predicted using elemental information without structural information [160]. For thermal conductivity prediction, Seko *et al*. showed that elemental descriptors containing atomic numbers, group, mass, radius, and ionization energy, and electron affinity along with unit cell volume and atomic density is enough to predict lattice thermal conductivity, without any further specification of the atomic structure of the unit cell [161]. Instead of using combined elemental and structural representations, direct use of vibrational properties as descriptors could also result in regression models with high fidelity, because vibrational properties have a more direct causality correlation with thermal conductivity compared with structural or elemental properties like space groups and ionic radius. Juneja *et al.* used vibrational properties including maximum phonon frequency, Grüneisen parameter, average atomic mass, and volume of unit cells as descriptors to build a GPR-based predictive model that can outperform the Slack relation by an order of magnitude [162]. Seko *et al.* also showed that prediction accuracy could be greatly enhanced by including the vibrational density of states (vDOS) as a descriptor, yet vDOS data is not always available in material databases [161].

While feature selection can be performed based on researchers' understanding of thermal transport physics, it is also possible to identify significant features in a data-driven manner without any pre-knowledge. Chen *et al*. investigated the relevant descriptors for high-throughput prediction of lattice thermal conductivity in inorganic materials through recursive feature elimination using random forest [163], with feature selection strategy discussed in Section 2.3.2. They divide the database equally into five subsets using the Bootstrap method, and recursively select one of the subsets as a training dataset and the rest as the validation datasets. For each step of training, correlation with thermal conductivity is computed for all the features and the least important features are removed from descriptor vectors. Finally, 29 relevant descriptors have been identified from the total 63 features for high-throughput prediction of thermal conductivity as listed in Figure 15a. Figure 15b shows the correlation map between the thermal conductivity and the five most important features, including bulk modulus, density, averaged sound speed (square root of bulk modulus divided by density), average bond length, and mean atomic mass. The general trend of increased thermal conductivity with increasing bulk modulus and sound speed is consistent with the physical intuitions. In addition, such analysis captured the trend of decreased thermal conductivity with heavier atomic masses and larger bond lengths, qualitatively agreeing with the Slack model. Nevertheless, such correlations cannot be simplified as power-law dependences.

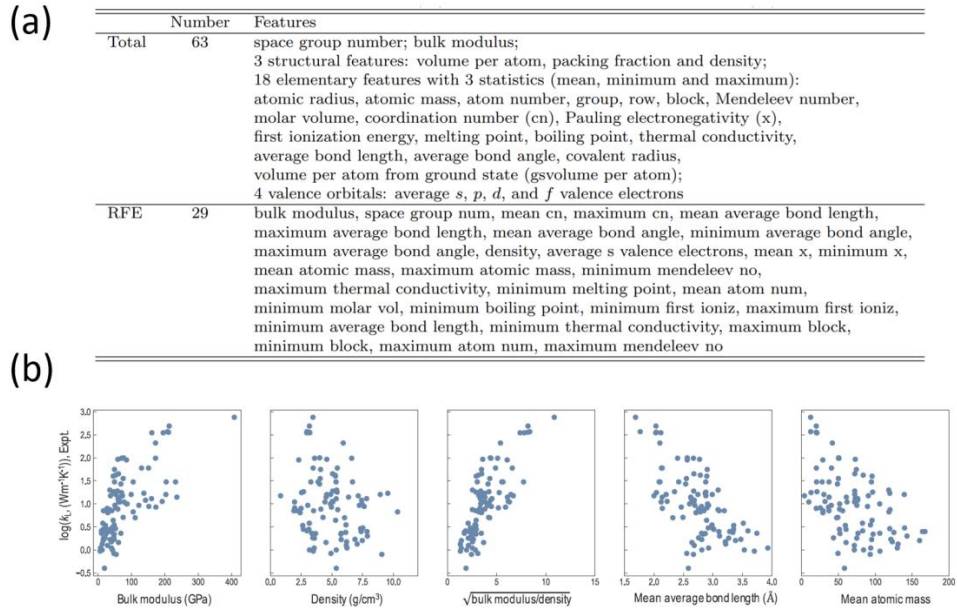

Figure 15. (a) List of possible features that can be used as descriptors for thermal conductivity prediction and the selected 29 most relevant features using recursive feature elimination (RFE). (b) Correlation map between thermal conductivity and five most relevant material properties. Reproduced from ref. [163], Copyright 2019 by Elsevier.

For the selection of ML algorithms for high throughput prediction of thermal conductivity, it is crucial to realize that the size of available training data is usually limited, especially after the pre-screening process. Although highly versatile nonlinear models such as deep neural networks have shown better performance compared with simpler algorithms like KRR when the dataset is large [164], these models might not perform well when the dataset is limited as it could result in overfitting problems for complex machine learning algorithms [71]. Random forest is one of the most popular choices when dealing with a relatively small size of the dataset. Carrete *et al*. showed that thermal conductivity predicted by random forest model tends to have centralized distribution around a mean value, which prevents unphysically high or low values [55]. Another good choice of algorithm for studying thermal transport properties is indeed GPR [161, 162], which is simple to implement, performs well with a relatively small training dataset, and explicitly computes uncertainty (variance) of the prediction. As shown by Seko *et al.* [161] [162], variances can give extra information on the maximum probability of finding material with better target thermal conductivity than the dataset. For discoverying low thermal conductivity materials, a quantity called "Z-score" is defined based on variance:

$$Z(\boldsymbol{q}^*) = \frac{f(\boldsymbol{q}^*) - f_{best}}{\sqrt{var[f(\boldsymbol{q}^*)]}} \tag{22}$$

where $f(\boldsymbol{q}^*) = -\ln[k_L(\boldsymbol{q}^*)]$ where $k_L(\boldsymbol{q}^*)$ is the predicted lattice thermal conductivity of a material labeled by the descriptor $\boldsymbol{q}^*$, $f_{best}$ denotes the function value with the lowest thermal conductivity among training dataset, and $var[f(\boldsymbol{q}^*)]$ is the variance of the predicted value at $\boldsymbol{x}^*$. Since $f$ decreases with $k_L$, a positive $Z$ score means that the predicted material could have lower thermal conductivity than the materials in the training dataset. The $\sqrt{var[f(\boldsymbol{q}^*)]}$ in the denominator would give a higher magnitude of $Z$-score for predictions with less uncertainty. Therefore, a large positive value of $Z$-score means that it is very promising to observe low thermal conductivity in the material labeled by the descriptor $\boldsymbol{q}^*$. Figure 16 shows the dependence of $Z$-score for lattice thermal conductivity for 54779 compounds in the dataset, along with the volume of unit cell $V$ and density of the material $\rho$. The fact that $Z$-score does not show clear correlation with $V$ and $\rho$ indicates that one cannot accurately predict thermal conductivity without elemental descriptors. The distribution of $Z$-score among elements could provide insights for selecting elements when looking for compounds with low or high thermal conductivity. For example, high $Z$-scores are found with heavy elements like I, Cs, and Pb, indicating that the compounds with these elements tend to show low thermal conductivity. On the other hand, $Z$-scores are widely distributed for light elements like Li, O, N, and F, implying that the presence of such elements has little or no effect on lowering the thermal conductivity. In particular, light elements like Be and B have a large negative Z-score, indicating the compounds with these light elements probably have high thermal conductivity.

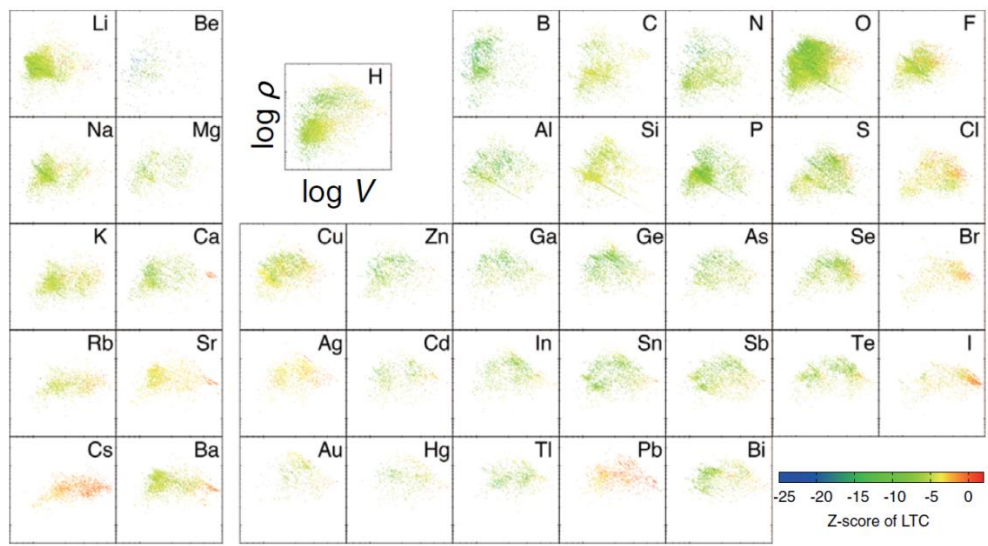

Figure 16. Dependence of Z-score as a function of the density $\rho$ and volume $V$ for each element. Reproduced from ref. [161], Copyright 2015 by American Physical Society (APS).

Transfer learning is another method often used for dealing with limited data. Recently, Ju *et al.* performed transfer learning for high-throughput prediction of thermal conductivity of diamond-like cubic crystals.[165] Transfer learning relies on the fact that many material properties are physically correlated, for example, phonon scattering phase space and thermal conductivity. Since the computation of scattering phase space only involves examining phonon frequency and momentum, it is much easier to evaluate compared with anharmonic phonon properties such as mean free paths. As a result, mapping the prediction of thermal conductivity to the prediction of scattering phase space can significantly enhance the size of available training data, as shown in Figure 17a. Therefore, one can first train a neural network model for predicting scattering phase space with a bigger dataset. To correlate scattering phase space and lattice thermal conductivity, one can couple the pre-trained neural network with another ML model which performs better, such as random forest, with a smaller training dataset for thermal conductivity. Figure 17b shows that ordinary ML has poor predictive power for a validation dataset with 14 diamond-like crystals, but when scattering phase space is included for transfer learning, the model has a much higher prediction accuracy. Although there still exist outliers with large deviations of thermal conductivity in Figure 17b, the accuracy of the transfer learning model is expected to improve with the increasing size of thermal conductivity databases and the computation power.

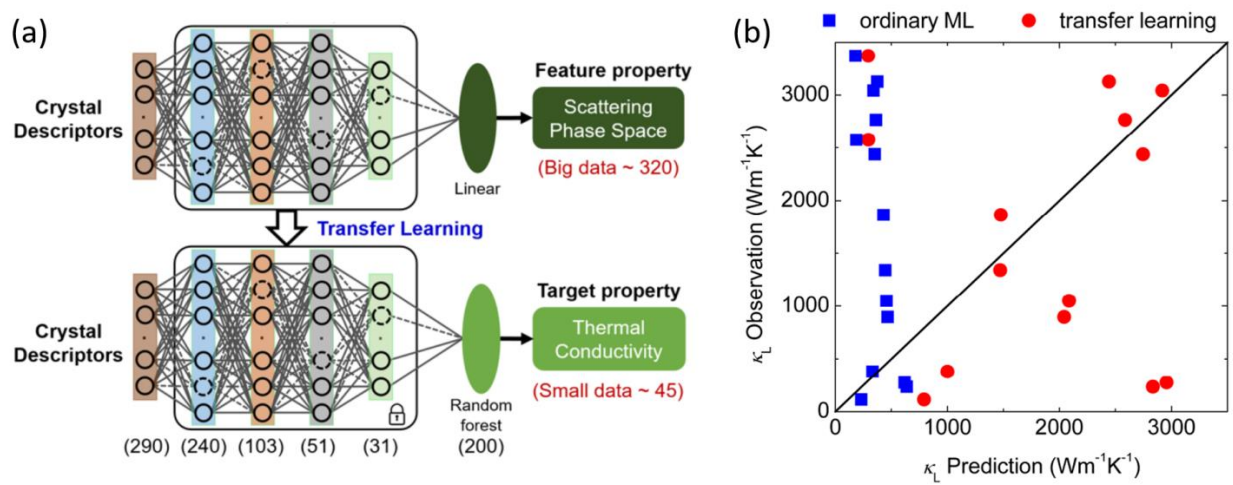

Figure 17. (a) Schematic of transfer learning by coupling neural networks (NN) and random forest. Numbers in the bracket indicate the number of neurons used in NN or the number of trees in the random forest. (b) Comparison between ordinary ML and transferred learning validated over 14 diamond-like crystals. Reproduced from ref. [165], Copyright 2019 by American Physical Society.

Using a similar transfer learning technique, Wu *et al.* performed high-throughput prediction for thermal conductivity of polymers [166]. For polymers, the size of data also varies across different properties and available data for thermal conductivity is indeed very limited. For example, the PoLyInfo library [167] recorded glass transition temperature and melting temperature for 5917 and 3234 unique homopolymers, respectively, whereas thermal conductivity values of only 28 homopolymers are available and only for near room temperature (10-35 °C). To deal with this challenge, the authors proposed the training target from thermal conductivity prediction to modeling glass-transition temperature and melting temperature with sufficient data available. High glass-transition temperature usually indicates rigid structures and probably high thermal conductivity as shown in Figure 18a. Figure 18b-c shows the performance of the pre-trained NN for predicting glass transition temperature and melting temperatures and the thermal conductivity prediction is compared in Figure 18c-d. Clearly, accurate predictions can be achieved when the transfer learning technique is used with correlation coefficient $R \sim 0.73$ for predicting values (prediction) and reference data (observation), while the $R$ for direct learning is as low as $-0.46$. The authors further utilized a molecular structure optimization algorithm to generate hypothetical polymer structures and applied the transfer learning model for predicting the thermal conductivity values of these polymers. They identified 24 novel polymers with possible high thermal conductivity, among which three have been experimentally synthesized with the highest thermal conductivity reaching 0.4 W/mK [166]. Comparing with polymers with typical thermal conductivity in the range of 0.1~0.3 W/mK [168], this work provided a preliminary demonstration of machine-learning enabled material discovery for macromolecules with optimal thermal transport properties.

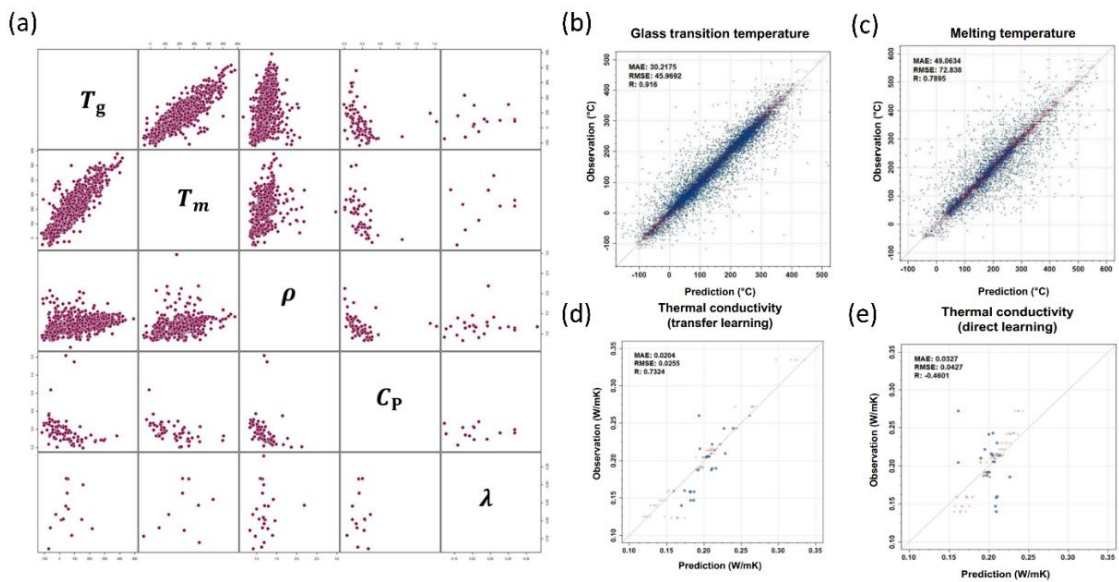

Figure 18. (a) Scatterplot matrix that summarizes the correlation map of the five polymeric properties including glass transition temperature $T_g$, melting temperature $T_m$, density $\rho$, specific heat $C_P$ and thermal conductivity $\lambda$. Cross-validation of (b) $T_g$ and (c) $T_m$ for the pre-trained model. (d-e) Cross-validation for polymer thermal conductivity through (d) transfer learning and (e) direct learning. Red dots indicate training data and blue dots indicate validation data. Mean absolute error (MAE) and root-mean-squared error (RMSE) and correlation coefficient ($R$) evaluated upon validation dataset are shown in the figure. This figure is reproduced from ref. [166], Copyright 2019 by Springer Nature.

Recently, Zhu et al.[169] combined transfer learning with convolution neural networks to reconcile the dilemma between data quality and data size.[52] Despite the large number (~$10^5$) of data entries inside ICSD database [52], the number of materials that have thermal conductivity data is only on the order of a few hundred with the majority of the data obtained from high throughput computations. However, the credibility of high-throughput predictions is also questionable due to the coarse convergence thresholds used in the DFT simulations in exchange for speed. That said, if only materials with reliable thermal conductivity values in the literature are considered, the available data size is even smaller. To solve this problem, the authors first pre-trained a convolution neural network model based on available computational data of thermal conductivities ($k_C$) from the database TE Design Lab.[54] To further account for the difference between larger online data $k_C$ and the literature experimental values ($k_{exp}$), the authors added one extra hidden layer to the neural network before the output layer. Training of the last layer is performed upon the experimental literature data $k_{exp}$, while keeping the parameters of the previous pre-trained layers fixed. Using a smaller set of experimental data with higher quality, such a "transfer learning" technique is able to improve the accuracy of the ML model pre-trained upon a larger database but with limited quality (Figure 19a). Using such transfer learned convolution neural networks, the authors have predicted the thermal conductivity of all ICSD entries with improved accuracy compared with other ML models (Figure 19b). The authors have used the trained transfer learning model to identify rare-earth chalcogenides as promising thermoelectric materials with low thermal conductivities 0.5 ~ 0.6 W/mK at 973 K (Figure 19c) [169]. Excitingly this has been experimentally validated with a promising thermoelectric material family of Bi-doped rare-earth chalcogenides at high temperatures (Figure 19d).

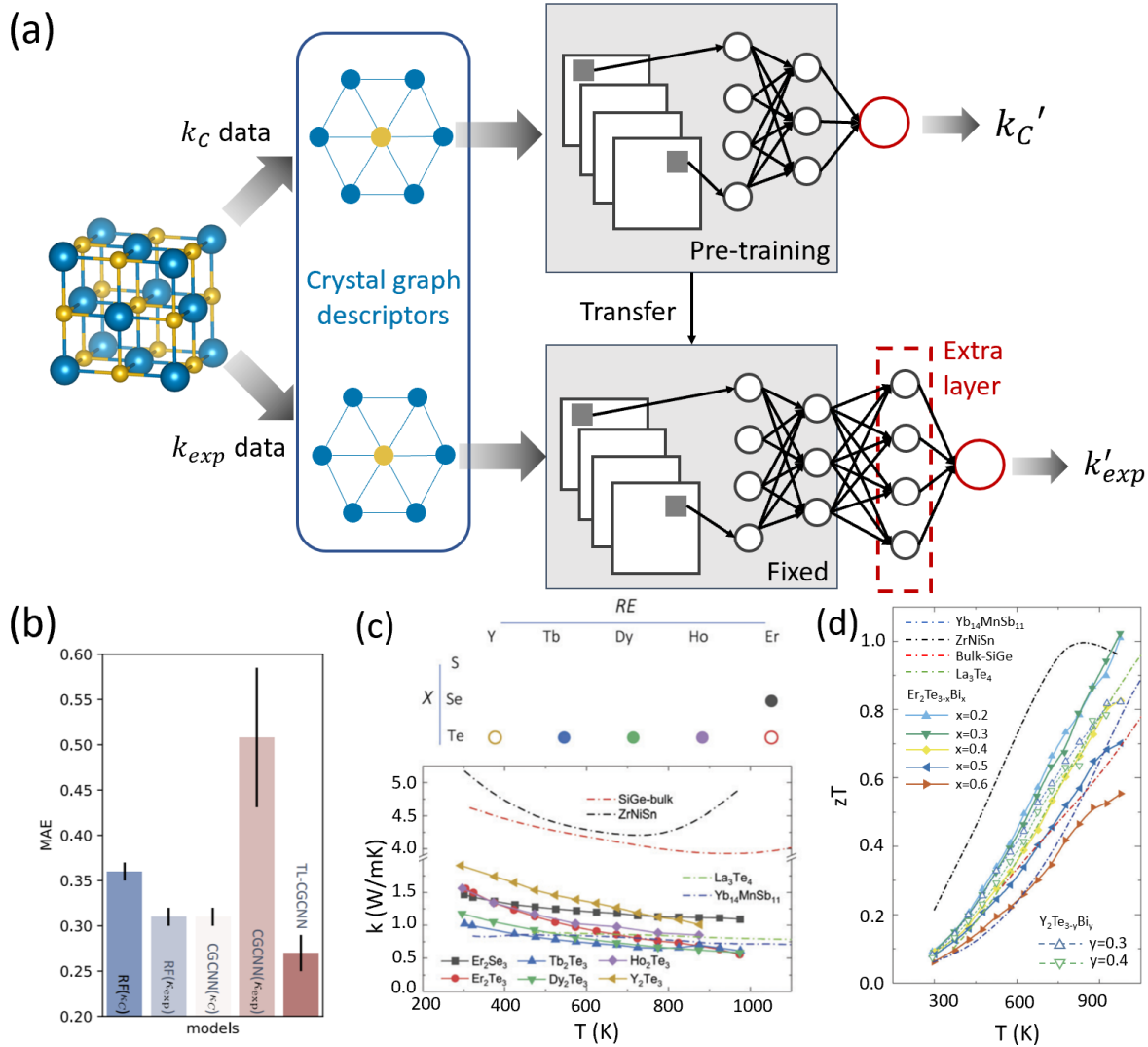

Figure 19. (a) Schematic of transfer learned crystal graph convolution neural network (TL-CGCNN). First, crystal graphs convert the crystalline structures to descriptors[83], which is a form of topological representation. Then the convolution neural network is pre-trained using computational data $k_C$. The pretrained convolution neural network gives prediction values $k_C'$. To correct the inconsistency between $k_C$ and experimental data $k_{exp}$, an extra layer is added in front of the output layer. This extra layer is trained with experimental data $k_{exp}$ while keeping the previous layer fixed. (b) Mean absolute error (MAE) of logarithmic thermal conductivity $lg(k)$ between different models: random forests trained using computational data RF($k_C$) and experimental data RF($k_{exp}$); crystal graph convolution neural networks (CGCNN) trained using $k_C$ and $k_{exp}$, denoted as CGCNN($k_C$) and CGCNN($k_{exp}$), and TL-CGCNN. (c) The experimental validation of low thermal conductivity of rare earth calcogenides (REX). The open symbols in the top panel indicate the materials not yet characterized while the filled symbols indicate those already validated by experiments. (d) Thermoelectric figure of merit ($zT$) of REX alloys. (b-e) is reproduced from ref.[169], Copyright 2021 by Royal Society of Chemistry (RSC).

## 3.3. Machine-learning-assisted design of atomic-architectured materials

The very first demonstration of applying ML-assisted design of the nanostructured material with optimal thermal transport property is by Ju *et al.* [170] The authors utilized the Bayesian optimization (BO) technique to identify nanostructured interfaces that yields minimum or maximum thermal conductance between Si and Ge, as shown in Figure 20a. Figure 20b shows the workflow for the structural design using BO technique. First, several SL structures are randomly constructed and the corresponding thermal conductance values were calculated using atomistic Green's function [149, 171, 172], which is then used as the training dataset $\mathcal{D}$ for a GPR prediction model. This prediction model is then used to search the next sampling point through maximizing or minimizing the "expected improvement" (EI) of thermal conductance:

$$EI(\boldsymbol{q}) = \max(f(\boldsymbol{q}) - f(\boldsymbol{q}_{opt}), 0) \qquad (23)$$

where $f$ is the function that predicts thermal conductance taking certain structure labeled by the vector $\boldsymbol{q}$, and $\boldsymbol{q}_{opt}$ labels the structure with the highest or lowest thermal conductance in the training dataset. The descriptor $\boldsymbol{q}$ is a binary vector containing zero and one, where zero represents a Si atom and one represents a Ge atom as shown in Figure 20a. The sampled structure is again evaluated by the atomistic Green's function for thermal conductance and added into the training dataset to update the prediction model. Such a process is iterated until the EI of thermal conductance approaches zero. Through the BO technique, Ju *et al.* identified the superlattice structures with both the maximum and the minimum thermal conductance as shown in Figure 20c. For the structure with the highest thermal conductance, there is a continuous path of Si or Ge atoms, which agrees well with the physical intuition that the continuous path provides a heat conduction channel. Whereas for the minimum thermal conductance, the superlattice involves an aperiodic arrangement of layers, which restrains the constructive interference of lattice waves [170]. However, the BO approach is only suitable for small size (hundreds or thousands) of candidate structures, which limits the length scale of interface structures that can be designed. Ju *et al.* further utilized a more efficient decision tree-based Monte Carlo tree search (MCTS) technique for larger simulation cells, with details presented in ref. [173].

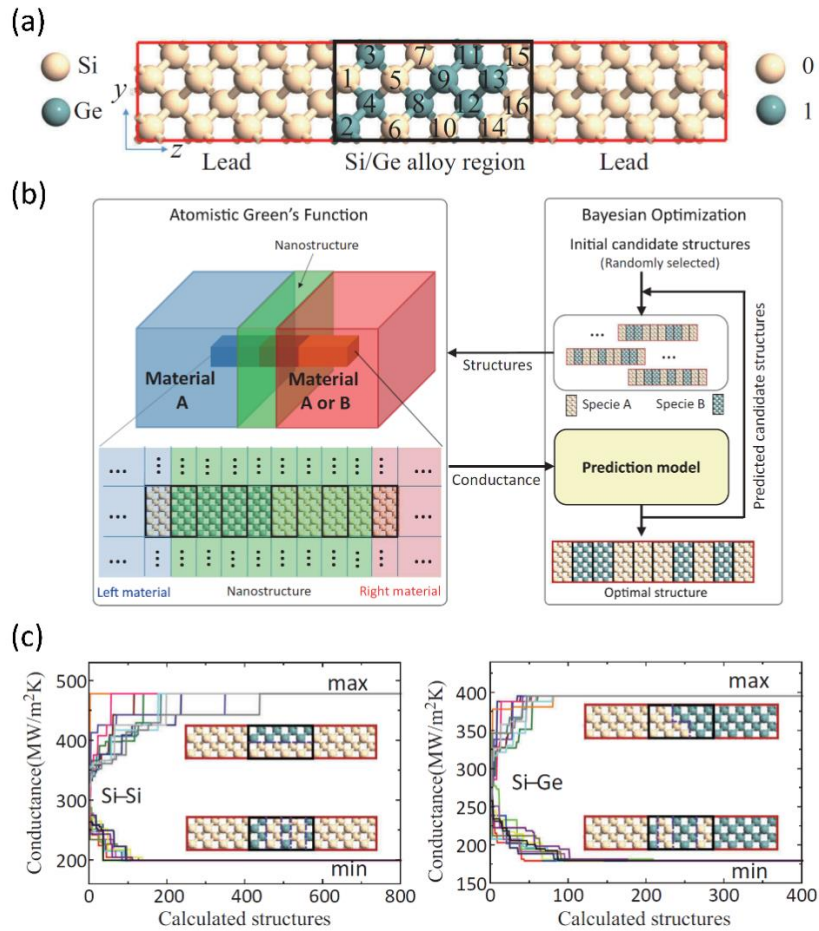

Figure 20. (a) Schematic of nanostructured Si-Ge interface, (b) Workflow of using Bayesian optimization for designing nanostructured interface with optimal thermal conductance. (c) Ten optimization runs with different initial training structures sandwiched between Si-Si and Si-Ge. The insets indicate the optimal structures with maximum or minimum conductance. Reproduced from ref. [170], Copyright 2018 by American Physical Society (APS).

Chakraborty *et al.* further investigated the key structural descriptors to minimize the thermal conductance of binary aperiodic superlattices and trained a neural network using the data from molecular dynamics simulations[174]. Intuitively, designing aperiodic superlattices involves parameters capturing the randomness of the layer thickness and superlattice period, therefore the standard deviation of the period seems to be a reasonable choice:

$$\delta = \sqrt{\frac{\sum_{i=2}^{N}\left[(d_{A,i} + d_{B,i}) - \langle d_A + d_B \rangle\right]^2}{N}} \quad (24)$$

where $d_{A,i}$ and $d_{B,i}$ are the thickness of material A and B in the binary aperiodic superlattice in the $i$-th period, $\langle d_A + d_B \rangle$ is the mean thickness of the "unit cell" containing both material A and B, and $N$ is the number of periods. However, such standard deviation failed to distinguish the permutation order of the thickness of different layers. For example, given a series of thickness $d_{A,i}$ and $d_{B,i}$, one can construct a gradient multilayered (GML) structure of superlattices by stacking the layers from small to large thickness. Such superlattice would have the same standard deviation $\delta$ compared with random multilayer (RML) structures with a random distribution of thickness and period (Figure 21a). As a result, the standard deviation fails to distinguish GML and RML as shown in Figure 21b, where the data clusters of GML and RML are overlapping along the dimension of $\delta$. To solve this issue, the authors defined randomness based on difference of period between neighboring unit cells, such that the stacking order is taken into account. The periodicity randomness can therefore be defined as:

$$R_p = \sqrt{\frac{\sum_{i=2}^{N}\left[(d_{A,i} + d_{B,i}) - (d_{A,i-1} + d_{B,i-1})\right]^2}{N}} \quad (25)$$

where $d_{A,i} + d_{B,i}$ is the period of the $i$-th unit cell. In addition to the periodicity randomness, thickness randomness of the binary superlattice is also defined:

$$R_d = \sqrt{\frac{\sum_{i=2}^{N}\left[(d_{A,i} - d_{A,i-1})^2 + (d_{B,i} - d_{B,i-1})^2\right]}{N}} \tag{26}$$

Figure 21c-d clearly showed that RML tend to have a higher $R_p$ and $R_d$ compared with GML, indicating that they are relevant descriptors for structural randomness. This work provided insights into quantifying the structural randomness for minimizing thermal conductance, which could be important for future combinations of ML-based prediction and optimization algorithms such as genetic algorithms [175] for larger-scale structural design beyond the atomistic level.

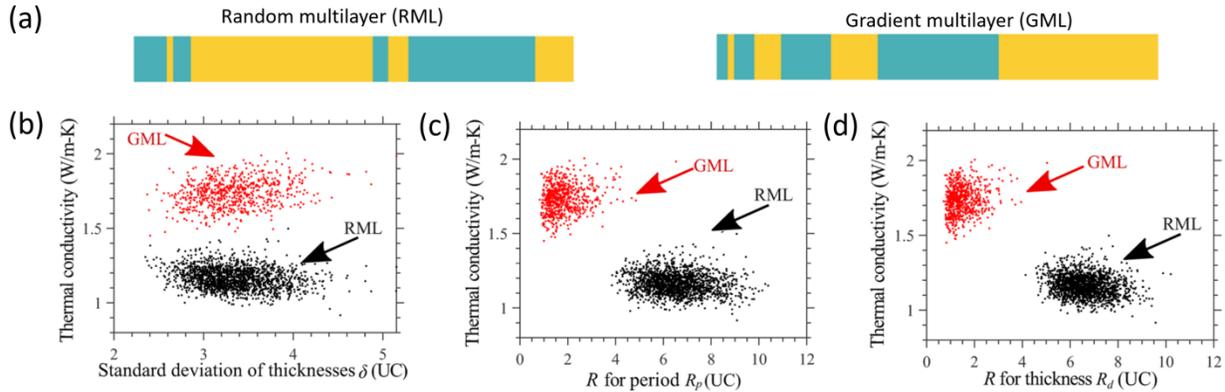

Figure 21. (a) Schematic of aperiodic superlattices with random multilayer (RML) geometry and gradient multilayer (GML) geometry with increased layer thickness. (b-d) Clustering of thermal conductivity data with respect to (b) standard deviation of unit cell thickness (c) randomness ($R$) of period, and (c) randomness of layer thickness. This figure is replotted based on ref. [174]. Copyright 2020 by American Chemical Society (ACS).

In addition to superlattices, Yamawaki *et al.* [176] designed the optimal distribution of antidots in graphene nanoribbons for optimal thermoelectric performances using the BO algorithm, as shown in Figure 22a-b. It is challenging to identify the optimal distribution of antidots due to the coupled electron and phonon transport. If the antidots are arranged periodically, an electronic bandgap would be opened which enhances the Seebeck coefficient. However, such periodicity would result in the interference effect of lattice waves, which is not desirable for minimizing thermal conductance. On the other hand, random distribution of antidots could induce phonon localization but would suppress electronic conductivity. The authors showed that BO can solve this design problem much more efficiently compared with random searches, as shown in Figure 22c. The optimal structure obtained from BO indeed involves an aperiodic array of antidots, with a nearly tenfold increase in the ZT factor as shown in Figure 22d.

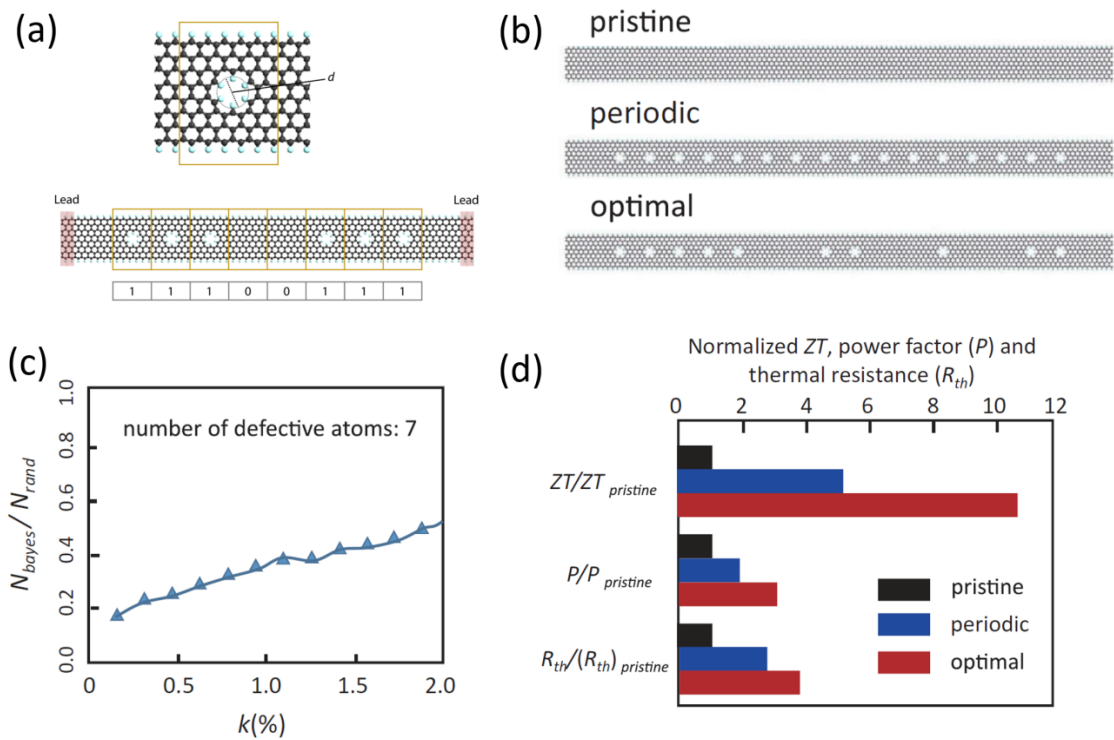

Figure 22. (a) Schematic of antidote distribution in graphene nanoribbons described by a binary vector. (b) illustration of a pristine graphene nanoribbon, a nanoribbon with periodic antidots, and a nanoribbon with optimal distribution of antidots (c) The ratio between the number of computation trials using Bayes search and random trial until identifying a structure with top $k\%$ of ZT. (d) ZT, power factor, and thermal resistance of graphene with periodic and optimal distribution of antidots, normalized by pristine graphene nanoribbons. Reproduced from ref. [173] and [176]. Copyright 2018 by AAAS and 2019 by Taylor and Francis Group.

Interestingly, unexpected regimes of thermal transport can also be discovered using ML-based optimization algorithms. For example, Bao and Ruan [177] recently found that introducing disordered patterning of pores does not always decrease the thermal conductivity. They identified a structure with a periodic arrangement of pores having higher thermal conductivity than pristine graphene. Figure 23a shows the optimization workflow, where the authors use a BTE model assuming gray approximation of phonon lifetime for estimating thermal conductivity and implement the genetic algorithm for maximizing the thermal conductivity. For each candidate structure selected from the genetic algorithm, more detailed nonequilibrium molecular dynamics (NEMD) simulations are performed to confirm whether the thermal conductivity of the aperiodic patterning has higher thermal conductivity than periodic arrangements. Through the genetic algorithm-based search, the authors managed to identify two disordered porous structures (Figure 23b) with thermal conductivity even higher than the periodic counterpart, while manual search failed to identify these special cases even with orders of high magnitude more computer hours. Analyses showed that high thermal conductivity structures only contain displaced pores along the horizontal or the vertical direction, which can form a channel for high heat fluxes (Figure 23c).

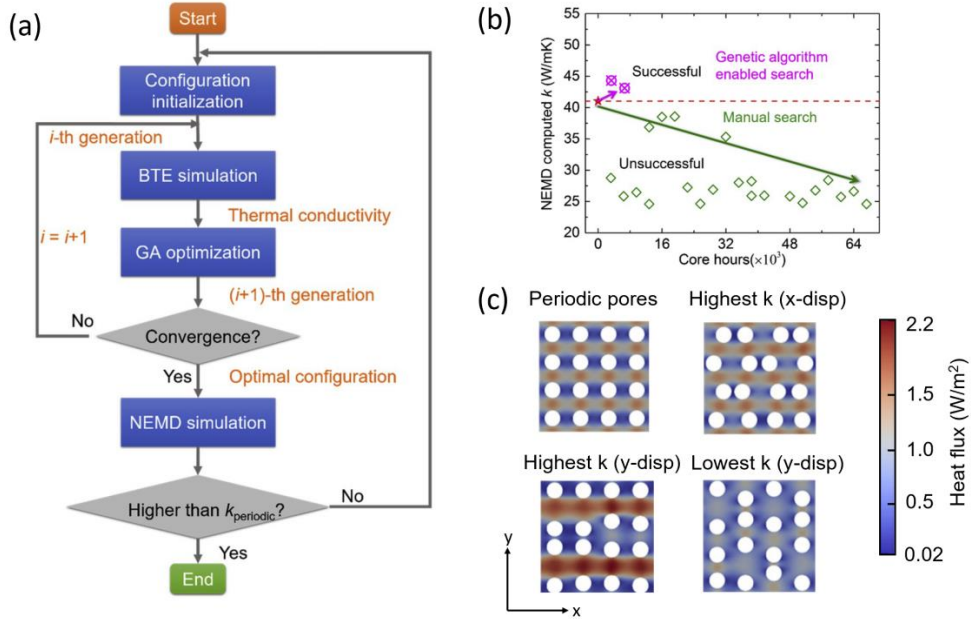

Figure 23. (a) The workflow of the genetic algorithm (GA)-enabled search process. BTE simulation is used for pre-screening and NEMD simulation is performed to validate the optimal configuration found by GA. (b) The comparison of the optimization output and the time cost between the genetic algorithm enabled search method (magenta) and manual search method (green). (c) distribution of heat flux for the periodic pore arrangements, the structure with aperiodic displacements in the x and y direction (labeled as x-disp and y-disp) identified by the GA optimization with high thermal conductivity, and the configuration with the disorder in y-direction identified by the GA optimization for low thermal conductivity. This Figure is reproduced from ref. [177], Copyright 2020 by Elsevier.

## 4. Summary and Outlook

This review provided a brief introduction to the concepts and the working principles of machine learning techniques and a comprehensive summary of early successes in applying machine learning techniques to study nanoscale thermal transport. In general, the state-of-the-art research using machine learning can be categorized into three ways as follows:

i) Machine learning regression techniques are used as fitting tools to build high-fidelity interatomic potentials, which are referred to as machine learning potentials. Such potentials free of rigid functional forms to describe interatomic bonding have shown excellent transfer-ability across different material systems including perfect crystals, crystals with defects, amorphous materials, and even different phases. Machine learning potential is especially powerful in modeling phonon and thermal transport properties for material systems that are challenging for DFT including crystals with defects and alloys, amorphous materials, and high-temperature materials.

ii) Another application of machine learning is for the high-throughput prediction and screening of materials, resonating well with the material genome initiative[178, 179] where several online material databases have been launched. Early successes of machine learning in thermal sciences have identified possible candidates in several material families such as perovskites [159], half-Heusler compounds [55], and rare-earth chalcogenides [169] with low thermal conductivity. The performance of machine learning models is significantly affected by the selection of relevant descriptors and machine learning algorithms that are suitable for the size of the available data.

iii) Finally, machine learning can be combined with simulation tools such as the first principles simulations and atomistic Green's function to identify the optimal structures for maximum or minimum thermal conductance.

Despite these early successes, there remain several technical challenges to be addressed for implementing machine learning techniques for studying nanoscale thermal transport.

i) Developing effective descriptors are always important and challenging. For building MLP, complete descriptors remain computationally expensive when compared with empirical potentials, especially for compounds consisting of multiple elements [106]. Computationally efficient descriptors for multiple element compounds, however, might not form a complete representation of atomic configurations. Such drawbacks can be complemented by using nonlinear machine learning models like neural networks, but at the risk of overfitting if the training dataset is not large enough [180]. For high-throughput screening purposes, different features might be inter-correlated or even irrelevant, therefore selecting the most relevant features is usually necessary to ensure the performance of ML models.

ii) Data availability is a challenge for studying thermal transport compared with other material properties. Therefore, one should be very careful in selecting machine-learning algorithms to prevent overfitting, instead of directly choosing the most versatile methods such as deep neural networks. Although first-principles phonon simulations are now routinely performed, the published data has not been well-organized into a database that is easily accessible.

iii) It is important to note that machine learning models do not predict well if the training data is inadequate. In particular, machine learning-based regression cannot be used as an extrapolation tool. When the root-mean-squared error (RSME) or mean absolute error (MAE) of a testing dataset is much larger than the training dataset, more data points especially unseen by the model should be included in the training set to improve the model. However, it could be computationally expensive to generate new data points for training. For example, it is computationally expensive to evaluate energy and forces by DFT for training MLP. A possible solution is through "active learning" [181]. Such a method actively selects training data points only in the region where the machine learning model tries to extrapolate beyond the training domain. Recently, such an "active learning" method is implemented in building MLPs [181, 182] and has been used for the prediction of thermal conductivity [183].

iv) The majority of the published work applying ML-driven material screening or structural design remains computational, and very few studies have experimental validation. The recent work by Hu and Iwamoto et al. [184] sets a good example of integrating ML-driven structural optimization, material synthesis, and thermal property measurements. ML optimization is first applied to obtain the aperiodic superlattice structure with minimal thermal conductivity. TDTR measurements were performed to characterize the superlattice after synthesized, showing excellent agreement of the measured value with ML prediction. Such integration of ML optimization with experimental characterization is encouraged in future studies.

v) Currently, the thermal science community has not realized the full power of machine learning techniques as it has not been directly related to high-throughput experiments. High-throughput measurements such as the pump-probe-based thermal property microscopy [185] could build thermal property images. Such measurements could potentially be combined with machine learning-based imaging processing techniques [186] to explore structural-property correlation and guide the design of nano-architectured materials and devices targeting optimal thermal transport. High-throughput measurements could also help to determine important descriptors. Recently, the thermal conductivities of high-entropy pyrochlore ceramics are studied using high-throughput measurements. As a result, high variance of bonding sizes and heavy atomic mass are identified as two important features promoting low thermal conductivity ceramics [187].

To conclude, the application of machine learning techniques in nanoscale thermal transport is still in its infancy and is expected to play a more pivotal role. This review serves as an important tutorial for researchers who are interested in applying machine learning techniques for studying thermal conductivity.

## Acknowledgment

R.Y. acknowledges the support from the National Natural Science Foundation of China (NSFC) under Grant No. 52036002. X.Q. acknowledges the startup funding granted by Huazhong University of Science and Technology (HUST). The authors declare no conflict of interest.